\documentclass[12pt]{article}

\usepackage{times}
\usepackage{times, xcolor, graphicx, graphics, cite, amsmath, multicol, upgreek}
\usepackage[hidelinks]{hyperref}
\usepackage[pagewise, modulo]{lineno}
\usepackage[labelfont={bf}]{caption}
\usepackage{booktabs, multirow, tabularx}
\newcolumntype{Y}{>{\centering\arraybackslash}X}
\newcounter{lastnote}

\begin{document} 
	\sloppy

	\baselineskip24pt

	{\parindent0pt 
		
		\Huge{Dose and compositional dependence of irradiation-induced property change in FeCr} 
		
		\bigskip
		
		\Large
		{Kay Song $^{\text{a}, \star}$, Dina Sheyfer $^{\text{b}}$, Kenichiro Mizohata $^{\text{c}}$,  Minyi Zhang $^{\text{d}}$, Wenjun Liu $^{\text{b}}$, Do\u{g}a G\"{u}rsoy $^{\text{b}}$, David Yang $^{\text{e}}$, Ivan Tolkachev $^{\text{a}}$, Hongbing Yu $^{\text{f}}$, David E J Armstrong $^{\text{d}}$, Felix Hofmann $^{\text{a}, \dagger}$}\\

		\large{$^{\text{a}}$ Department of Engineering Science, University of Oxford, Parks Road, Oxford, OX1 3PJ, UK} \\
		\large{$^{\text{b}}$ Advanced Photon Source, Argonne National Laboratory, 9700 South Cass Avenue, Argonne, IL 60439, USA} \\
		\large{$^{\text{c}}$ Department of Physics, University of Helsinki, P.O. Box 43, 00014 Helsinki, Finland} \\
		\large{$^{\text{d}}$ Department of Materials, University of Oxford, Parks Road, Oxford, OX1 3PH, UK}\\
		\large{$^{\text{e}}$ Condensed Matter Physics and Materials Science Department, Brookhaven National Laboratory, Upton, NY 11973-5000, USA} \\
		\large{$^{\text{f}}$ Canadian Nuclear Laboratories, Chalk River, ON K0J 1J0, Canada} \\
		
		\bigskip
		\large{$^{\star}$Corresponding author email: kay.song@eng.ox.ac.uk} \\
		\large{$^{\dagger}$felix.hofmann@eng.ox.ac.uk}
		
		\bigskip
		
		\begin{multicols}{2}
		\large{ORCID:}\\
		\normalsize{Kay Song: 0000-0001-8011-3862 \\ Dina Sheyfer: 0000-0003-4189-3899 \\ Kenichiro Mizohata: 0000-0003-1703-2247 \\ Minyi Zhang: 0000-0003-4131-9774 \\ Wenjun Liu: 0000-0001-9072-5379 \\ Do\u{g}a G\"{u}rsoy: 0000-0002-0080-8101 \\ \\ David Yang: 0000-0002-1062-7371 \\ Ivan Tolkachev: 0000-0001-8985-726X \\ Hongbing Yu: 0000-0002-6527-9677 \\ David E J Armstrong: 0000-0002-5067-5108 \\ Felix Hofmann: 0000-0001-6111-339X}
		\end{multicols}
	}
	
	\newpage
	\begin{abstract}
		Ferritic/martensitic steels will be used as structural components in next generation nuclear reactors. Their successful operation relies on an understanding of irradiation-induced defect behaviour in the material. In this study, Fe and FeCr alloys (3--12\%Cr) were irradiated with 20 MeV Fe-ions at 313 K to doses ranging between 0.00008 dpa to 6.0 dpa. This dose range covers six orders of magnitude, spanning low, transition and high dose regimes. Lattice strain and hardness in the irradiated material were characterised with micro-beam Laue X-ray diffraction and nanoindentation, respectively. 
		
		Irradiation hardening was observed even at very low doses (0.00008 dpa) and showed a monotonic increase with dose up to 6.0 dpa. Lattice strain measurements of samples at 0.0008 dpa allow the calculation of equivalent Frenkel pair densities and corrections to the Norgett-Robinson-Torrens (NRT) model for Fe and FeCr alloys at low dose. NRT efficiency for FeCr is 0.2, which agrees with literature values for high irradiation energy. Lattice strain increases up to 0.8 dpa and then decreases when the damage dose is further increased. The strains measured in this study are lower and peak at a larger dose than predicted by atomistic simulations. This difference can be explained by taking temperature and impurities into account. 
	\end{abstract}
	
	
	\newpage
	
	\begin{center}
		\small{\textbf{Graphical Abstract}}
		\begin{figure}[h!]
				\includegraphics[width=0.85\textwidth]{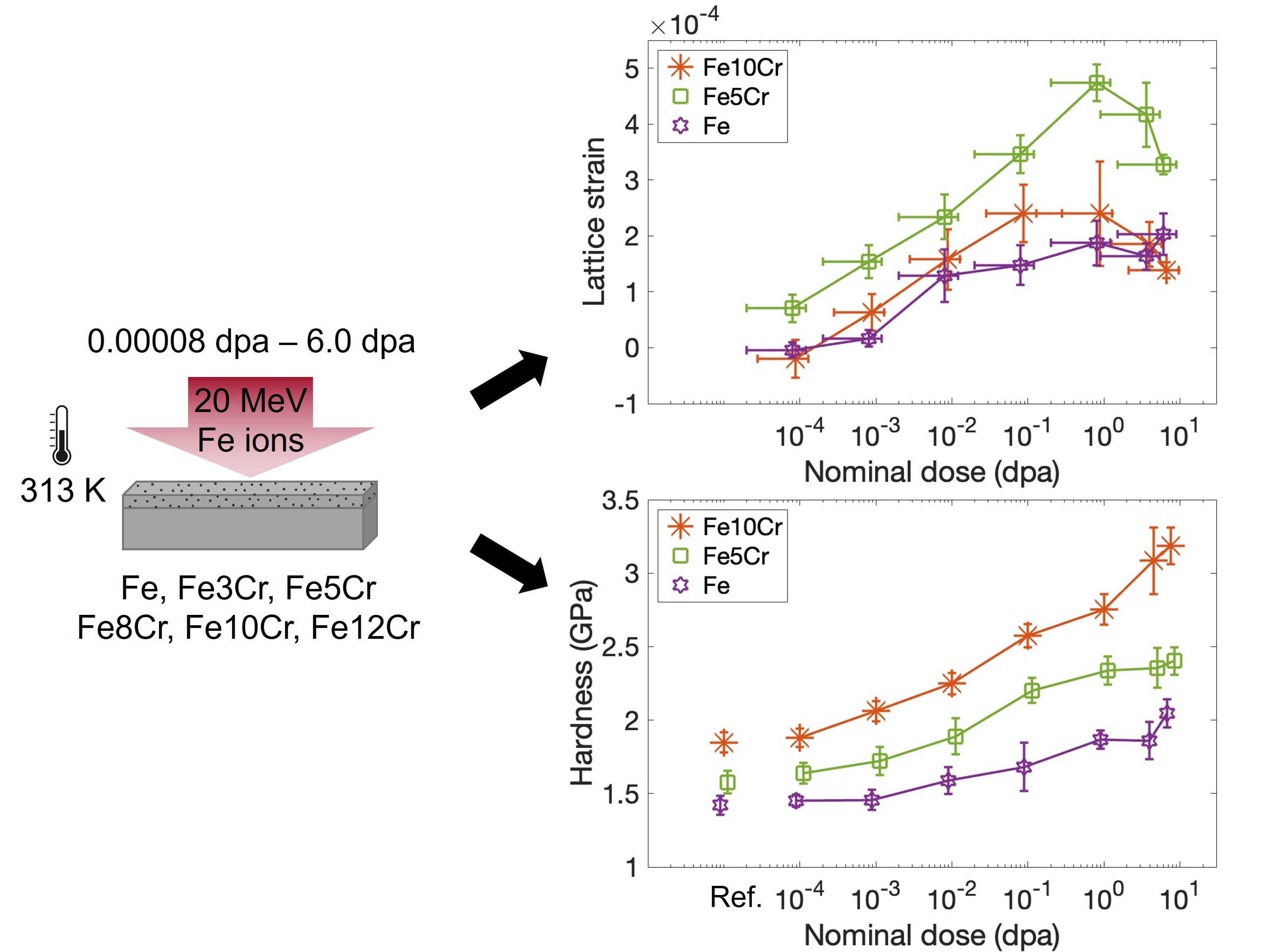}
				\centering
			\end{figure}
	
	\end{center}
	
	\newpage
	
		
		\section{Introduction}
		Reduced-activation ferritic/martensitic (RAFM) steels are leading candidate materials for structural components of fusion reactors, particularly for blanket/first-wall components \cite{Kurtz2019}. They possess superior resistance to radiation-induced swelling and better thermal properties than austenitic stainless steels \cite{Baluc2004}.
		
		To better understand the effect of irradiation on the microstructure and material properties of RAFM steels, iron-chromium (FeCr) binary alloys have often been studied as a model system \cite{Little1979, Kayano1988, Garner2000, Vortler2008, Bhattacharya2016}. The FeCr system can provide a mechanistic understanding of irradiation damage in ferritic/martensitic steels while removing many microstructural complexities. Consequently, it becomes more feasible to compare experimental and simulation data \cite{Malerba2008}. 
		
		A significant challenge in fully understanding the effects of nuclear fusion operations on structural steels is the lack of an existing power plant facility. While certain aspects of irradiation can be replicated with fission neutrons \cite{Zinkle2009, Knaster2016, Tanigawa2017} and ion irradiation \cite{Abromeit1994, Was2015, Harrison2019}, it is costly and time-intensive to reproduce the full range of conditions expected for structural materials in a nuclear fusion power plant. This necessitates the development of reliable simulation models ranging from first principles atomistic simulations to simulations that capture evolution and degradation at the component scale \cite{Zinkle2005, Malerba2008, Granberg2020, Gilbert2021}.
		
		Comparison to experimental data is crucial for the validation and optimisation of these models. However, several key gaps exist in the literature on FeCr regarding experimental studies:
		\begin{enumerate}
			\item \underline{Low-temperature ($<$ 573 K) irradiation effects:} The majority of experiments have been performed in the 623--773 K temperature window expected in-service for most of the structural components \cite{Bhattacharya2022}. Hardening following ion-irradiation has been observed at these temperatures, reaching saturation around 1--2 dpa, particularly for Cr content greater than 5\% \cite{Heintze2011, Hardie2013}. It is interesting to note that most of the data in the literature regarding hardening following neutron irradiation indicate hardness saturation beyond 5--10 dpa \cite{Bhattacharya2022}. This could be a result of different dose rates and primary knock-on atom energy spectra between neutron- and ion-irradiation \cite{Hardie2016}. Extensive transmission electron microscopy (TEM) observations have revealed the presence of $a\langle 100 \rangle$ and $\tfrac{a}{2}\langle 111 \rangle$ dislocation loops, with the former more dominant at temperatures closer to 773 K and the latter prevailing at 573 K or below \cite{Jenkins2009}. At high temperatures, the stability of dislocation loops is reduced \cite{Arakawa2004} and dislocation microstructures also undergo coarsening, both of which affect their distribution and density \cite{Horton1982}. 
			
			However, the effects of irradiation at low temperatures have not been extensively studied. Low temperatures are also of operational significance as reactor cooling components are expected to operate below 573 K \cite{Vaghela2021}. 
			Experimental data is required from irradiation in this temperature range, particularly at room temperature, in order to study and model the athermal effects of defect population and behaviour at lower temperatures.
			
			\item \underline{An extensive range of irradiation levels:} Reactor components in operation are expected to sustain irradiation exposure up to $\sim$100 dpa \cite{Gelles1995, Zinkle2009}.  Volumetric swelling, caused by the formation of voids at high doses ($>$ 1 dpa) and elevated temperatures ($>$ 673 K), has been extensively investigated for Fe and FeCr-based alloys \cite{Lin2021, Sencer2000}. At low doses ($<$ 0.1 dpa), volumetric swelling has been demonstrated to originate from the lattice strain of point defects \cite{Derlet2020}. The issue for reactor components is that even at low dose, lattice swelling can cause large stresses on the order of hundreds of MPa \cite{Dudarev2018a, Hofmann2015} and their non-uniform distribution due to irradiation will lead to macroscopic deformation. In FeCr, lattice strains are strongly dose- and composition-dependent \cite{Song2020}, but the parameter space explored thus far is limited.
			
			Studying material property changes at low doses ($\ll$ 1 dpa) is important for several reasons. Firstly, many irradiation-induced changes reach saturation at a certain dose \cite{Hardie2013, Mason2020, Armstrong2013}. Determining the dose threshold for saturation is important. Secondly certain modelling techniques, such as \textit{ab initio} and molecular dynamics \cite{Gilbert2021}, face limitations when trying to simulate a large range of exposures due to constraints in computational resources. Therefore, the availability of experimental results obtained at a range of, including very low, doses for comparison is needed. Thirdly, the distribution of defects and material property changes in a real-life reactor component will not be homogeneous. Predicting how these distribution gradients will affect material performance requires knowledge of material property changes across a wide range of damage levels. Finally, the damage microstructure of materials has been found to depend on pre-existing damage \cite{Gao1996, Sand2018}. As such, knowledge of the damage accumulation history and effects at lower doses is crucial.  
			\item \underline{The synergistic effect of dose and composition:} Though there are many studies of irradiation-induced effects in FeCr that explore a range of sample conditions \cite{Bhattacharya2016, Bhattacharya2022, Kurtz2019, Spatig2019}, the parameter space covered in each individual study (e.g. dose level, compositional variation, temperature, irradiating ion species) is generally limited \cite{Bhattacharya2016, Brimbal2013}. For example, the nanoindentation work of Heintze \textit{et al.} \cite{Heintze2011} only covers 2 doses (1 and 10 dpa) for Cr content ranging from 2.5\% to 12.5\% following ion irradiation at room temperature. There is of course a trade-off between examining specific phenomena in detail versus covering a broader parameter space. However, identifying synergistic effects, as well as broader parameter space effects, from comparing different studies is difficult due to variations in sample history and irradiation conditions. 
		\end{enumerate}
		
		To address these aforementioned gaps in literature, it is important to also consider which irradiation effects to focus on. TEM is one of the most common methods of studying irradiation-induced defect structures and populations \cite{Prokhodtseva2013, Yao2008, Jenkins2009}. TEM has provided many key insights on defect density, size distribution, and defect type. However, the lack of sensitivity of TEM to small defects ($<$ 1 nm) \cite{Zhou2006} makes it incomplete for the study of the full defect population, particularly at low doses where a majority of defects are below the detection threshold \cite{Yi2015}. Positron annihilation spectroscopy revealed a 2 order of magnitude discrepancy in defect cluster density compared to TEM measurements at doses as low as 10$^{-3}$ dpa \cite{Zinkle2006}. Measurements of material properties such as thermal diffusivity \cite{Reza2020} and lattice strain \cite{Song2020} have also revealed key insights into defect populations, while also showing significant discrepancies in defect density estimates compared to TEM studies at the same dose. The study of irradiation-induced changes in the mechanical properties of steels has also been useful to understand the effect of the damage microstructure \cite{Hosemann2009, Armstrong2015}. Therefore, lattice strain and hardness characterisation have been chosen for this study to obtain a broader understanding of defect populations in FeCr. Furthermore, key mechanistic insight can also be revealed by comparing and contrasting the evolution of different material properties.
		
		In this study, we present the characterisation of lattice strain and hardness changes induced by Fe-ion irradiation of Fe and FeCr alloys. We examine a range of doses from 0.00008 dpa to 6 dpa, spanning six orders of magnitude, for six compositions of FeCr binary alloys (0 $\leq$ Cr\% $\leq$ 12). We discuss key insights, at low dose, of defect production and retention rate, comparing with the commonly-used Norgett-Robinson-Torrens (NRT) model and microscopy results from the literature. The effect of dose and Cr concentration on defect mobility and clustering is explored by comparing experimental trends of hardness and lattice strain, as well as with simulations.

		\section{Materials and Methods}
		\subsection{Sample preparation and ion-implantation}
		
		The high-purity FeCr alloy materials used in this investigation were manufactured under the European Fusion Development Agreement (EFDA) programme (contract EFDA-06-1901). The chemical compositions and mean grain sizes of the as-delivered alloys are listed in Table \ref{tab:composition}. 
		
		\begin{table} [h!]
			\begin{center}

				\begin{tabularx}{\textwidth}{c *{5}{Y} c }
					\toprule
					Alloy	&	Cr (wt \%) 	&	C (wppm)	&	S (wppm)	&	O (wppm) 	&	N (wppm)	 & Mean grain	\\
					&  \multicolumn{5}{c}{(equivalent atomic content, at\% or appm, in brackets)} & size (\textmu m) \\
					\midrule
					\multirow{2}{*}{Fe}	&	$<$ 0.0002	&	4 	&	2 	&	4 	&	1 	&  \multirow{2}{*}{187 $\pm$ 150} \\
					& ($<$ 0.0002) & (18) & (3) & (14) & (4) & \\
					\addlinespace
					\hline
					\addlinespace
					\multirow{2}{*}{Fe3Cr}	&	3.05	&	4	&	3	&	6	&	2	& \multirow{2}{*}{160 $\pm$ 114} \\
					& (3.27)& (18) & (5) & (21) & (8) & \\
					\addlinespace
					\hline
					\addlinespace
					\multirow{2}{*}{Fe5Cr}	&	5.40	&	4	&	3	&	6	&	2	& \multirow{2}{*}{112 $\pm$ 60} \\
					& (5.78)  & (18) & (5) & (21) & (8) & \\
					\addlinespace
					\hline
					\addlinespace
					\multirow{2}{*}{Fe8Cr}	&	7.88	&	6 	&	2 &	10	&	2	& \multirow{2}{*}{86 $\pm$ 63} \\
					& (8.41) & (28) & (3) & (35) & (8) & \\
					\addlinespace
					\hline
					\addlinespace
					\multirow{2}{*}{Fe10Cr}	&	10.10	&	4	&	4	&	4	&	3 	& \multirow{2}{*}{98 $\pm$ 55} \\
					& (10.77) & (18) & (7) & (14) & (12) & \\
					\addlinespace
					\hline
					\addlinespace
					\multirow{2}{*}{Fe12Cr}	&	11.63	&	6	&	2	&	4	&	$<$ 10	& \multirow{2}{*}{281 $\pm$ 250} \\
					& (12.38) & (28) & (3) & (14) & ($<$ 40) & \\
					\addlinespace
					\bottomrule
				\end{tabularx}
			
				\caption{The chemical compositions of the FeCr alloys used in this study as measured by the manufacturer using glow discharge mass spectrometry \cite{Coze2007, Fraczkiewicz2011}. The manufacturer reported values in weight content, which have also been converted to atomic content as shown in the table. The mean grain size was determined from EBSD measurements, with $\pm$ 1 standard deviation shown here (see Appendix A).}
				\label{tab:composition}
			\end{center}
		\end{table}
	
		The alloys were produced by induction melting under an argon atmosphere, followed by hot-forging at 1273 K for Fe and 1423 K for all other compositions. The bars were subsequently cold-forged with a reduction ratio of 70\%. Heat treatment of 1 hour was conducted at 973 K for Fe, 1023 K for Fe3Cr, Fe5Cr, and Fe8Cr, 1073 K for Fe10Cr, and 1123 K for Fe12Cr. The recrystallised materials were then air-cooled \cite{Coze2007, Fraczkiewicz2011}. Surprisingly, for Fe8Cr, our electron backscatter diffraction (EBSD) characterisation of the materials revealed significant signs of cold work (intragranular misorientation) in the microstructure of the materials, unlike the other compositions (Appendix A). In addition, the measured average grain size of 86 $\pm$ 63 \textmu m differs noticeably from the manufacturer-quoted value of 320 \textmu m \cite{Fraczkiewicz2011}. As such, we suspect that the Fe8Cr raw material may not have been fully heat treated to the conditions reported by the manufacturer.
		
		The as-delivered samples were sectioned with a fast diamond saw into pieces of approximately 5 $\times$ 5 $\times$ 0.7 mm$^{3}$ in size. The polishing process consisted of mechanical grinding with SiC paper, followed by polishing with diamond suspension and colloidal silica. Finally, the samples were electropolished with 5\% perchloric acid in ethanol, with 15\% ethylene glycol monobutyl ether, at 293 K for 2--3 minutes using a Struers LectroPol-5. The voltage applied during electropolishing was 45 V for Fe and Fe3Cr, 35 V for Fe5Cr and Fe8Cr, and 30 V for Fe10Cr and Fe12Cr.
		
\begin{figure}[h!]
			\centering
			\includegraphics[width=0.6\textwidth]{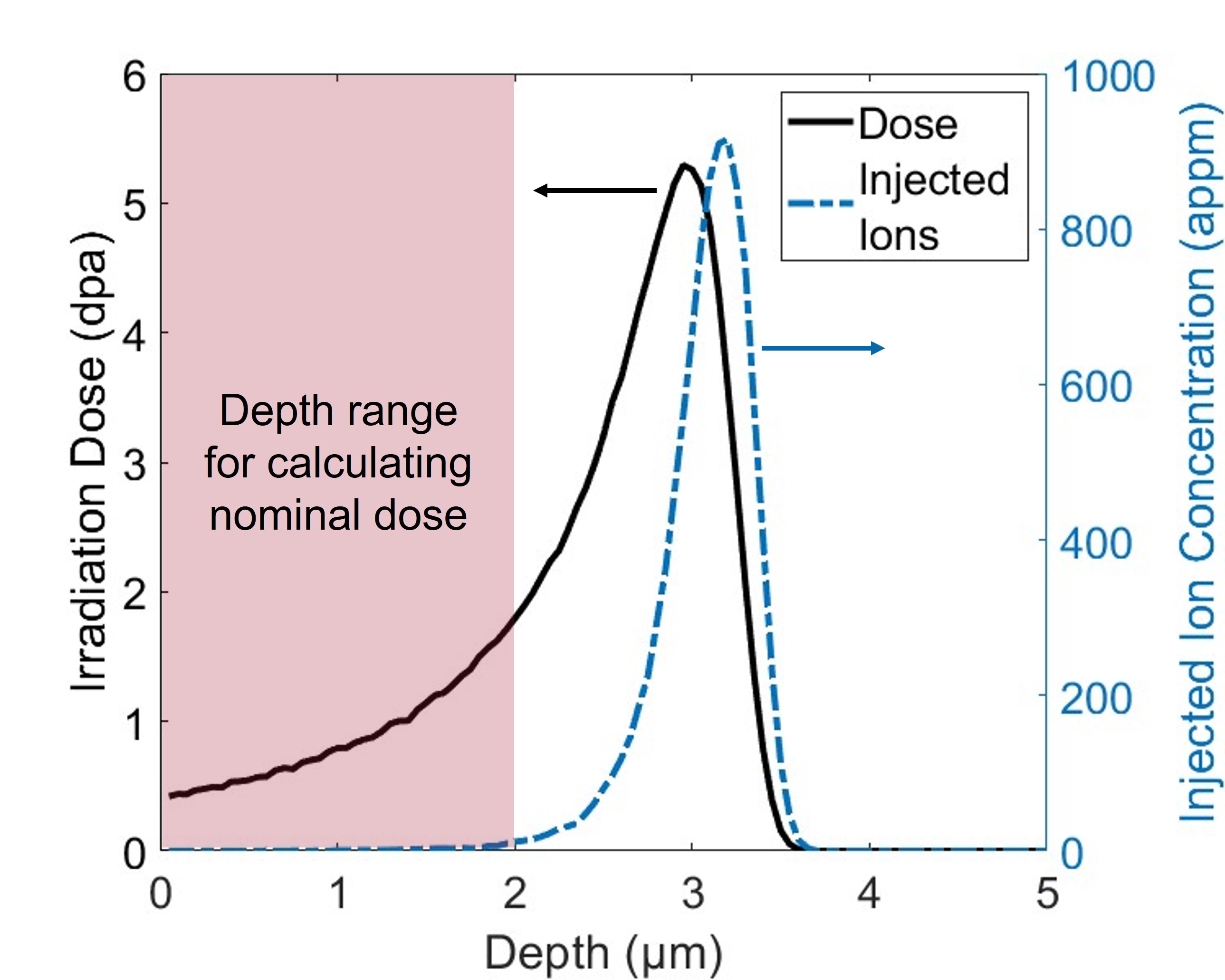}
			\caption{The calculated dose and injected ion profile from SRIM for 20 MeV Fe-ion irradiation on Fe target. The profile shown corresponds to a nominal dose of 0.8 dpa (average of the first 2 \textmu m of the dose profile). It is assumed the total profile of the dose and injected ions scale linearly with irradiation ion fluence. }
			\label{fig:dpa}
		\end{figure}

		An energy of 20 MeV was chosen for Fe-ion irradiation, as this produced a damage layer of 3.5 \textmu m thickness, allowing characterisation with X-ray diffraction and nanoindentation. Depth profiles of the damage and injected ion distribution (Figure \ref{fig:dpa}) were calculated with SRIM using the Quick K-P model \cite{Ziegler2010} with 20 MeV Fe ions on a Fe target with 40 eV displacement energy \cite{ASTME5212016} at normal incidence. The dpa profile was calculated using the vacancy.txt method described in \cite{Stoller2013, Agarwal2021}. In this study, the nominal dose of an irradiation profile refers to the average damage dose in the first 2 \textmu m of the sample, where the damage profile is relatively flat and the injected ion concentration is still low.
		
		Ion-implantation was performed at room temperature with Fe$^{4+}$ ions using the tandem accelerator at the Helsinki Accelerator Laboratory. The vacuum level inside the irradiation chamber was maintained at $8\times 10^{-7}$ mbar. The irradiation conditions are listed in Table \ref{tab:irradiation}. For each sample composition, up to 7 different nominal dose levels were produced (0.00008 dpa, 0.0008 dpa, 0.008 dpa, 0.08 dpa, 0.8 dpa, 3.6 dpa and 6.0 dpa). An unirradiated reference sample was also retained for each composition. To enhance readability, we will use scientific notation labelling for the lowest 4 doses in this study. 0.8 is used as the coefficient (e.g. 0.00008 dpa is written as 0.8E-4 dpa) to align the exponents in the notation with the exponents of the figures presented in the Results section. 
		
		A custom-built holder was designed for the irradiation process that allowed active temperature control using a combination of liquid nitrogen cooling and a cartridge heater. The samples were mounted on an aluminium block, and a thermocouple was positioned within 10 mm of the samples to monitor the temperature. For the irradiations in this investigation, the temperature was held constant at 313 K. Due to the spatial constraints of the sample holder, the samples had to be irradiated in two separate groups (Fe/Fe3Cr/Fe5Cr and Fe8Cr/Fe10Cr/Fe12Cr). The exception was for the 6.0 dpa irradiation where only Fe, Fe5Cr, Fe10Cr, and Fe12Cr were irradiated simultaneously due to time constraints.
		
		\begin{table}  [h!]
			\begin{center}
				\begin{tabular}{ cccccc }
					\toprule
					Dose name & Nominal dose	&	Total fluence	&	Nominal dose &	Flux 	&	Irradiation time	\\
					 & (dpa)	&	 (ions/cm$^{2}$)	&	rate (dpa/s)	&	(ions/cm$^{2}$/s)	&	(hour:min:sec)	\\
					\midrule
					0.8E-4 & 0.00008	&	5.30$\times 10^{11}$	&	3.54$\times 10^{-6}$	&	2.34$\times 10^{10}$	&	00:00:17	\\
					0.8E-3 & 0.0008	&	5.30$\times 10^{12}$	&	3.54$\times 10^{-6}$	&	2.34$\times 10^{10}$	&	00:03:12	\\
					0.8E-2 & 0.008	&	5.30$\times 10^{13}$	&	4.65$\times 10^{-5}$	&	3.08$\times 10^{11}$	&	00:02:51	\\
					0.8E-1 & 0.08	&	5.30$\times 10^{14}$	&	5.31$\times 10^{-5}$	&	3.52$\times 10^{11}$	&	00:25:06	\\
					0.8 & 0.8	&	5.30$\times 10^{15}$	&	2.78$\times 10^{-5}$	&	1.8$\times 10^{11}$	&	08:09:27	\\
					3.6 & 3.6	&	2.39$\times 10^{16}$	&	4.61$\times 10^{-5}$	&	3.05$\times 10^{11}$	&	21:42:00	\\
					6.0 & 6.0	&	3.98$\times 10^{16}$	&	4.50$\times 10^{-5}$	&	2.98$\times 10^{11}$	&	37:00:16	\\
					\bottomrule
				\end{tabular}
				\caption{Irradiation conditions for this study. The nominal dose is calculated from the average of the dose profile in the top 2 \textmu m below the surface, as shown in Figure \ref{fig:dpa}. The convention described for the dose name will be used for the rest of this paper.}
				\label{tab:irradiation}
			\end{center}
		\end{table}
				
		\subsection{Micro-beam Laue X-ray diffraction}
		Lattice strain, i.e. change in the atomic plane-spacing of the crystal, caused by the irradiation was measured using micro-beam Laue X-ray diffraction at the 34-ID-E beamline, Advanced Photon Source (Argonne National Laboratory, IL, USA). Depth-resolution in the sample measurements was obtained using Differential Aperture X-ray Microscopy (DAXM), which has been described in detail elsewhere \cite{Larson2002, Liu2004, Das2018a}. Briefly, the sample is mounted in 45$^{\circ}$ reflective geometry. An area detector (Perkin-Elmer, \#XRD 1621, pixel size 200 $\times$ 200 \textmu m$^{2}$) positioned above the sample records the Laue diffraction patterns. The key component of this technique is a platinum wire ($\sim$100 \textmu m in diameter, mounted on a silicon monocrystal) that is oriented perpendicular to the beam direction, and scanned parallel to the surface of the sample through the diffracted beams. By comparing the intensity of each detector pixel as the wire is moved along consecutive positions, and triangulating using the position of the incident beam and the wire edge, a profile of intensity as a function of sample depth can be obtained for the whole detector. This enables the reconstruction of diffraction patterns as a function of depth into the sample along the beam path (i.e. at 45$^{\circ}$ to the sample surface). By also scanning the incident X-ray beam energy, a Bragg peak can be fully measured in 3D reciprocal space and as a function of depth into the sample. Note that for the subsequent analysis presented in this study, all profiles have been converted into a function of depth perpendicular to the sample surface.
		
		The X-ray beam size at the sample surface was 190 $\times$ 360 nm$^{2}$, and the estimated depth resolution was $\sim$1 \textmu m. For each sample, a minimum of 3 points were measured on grains within 10$^{\circ}$ of $\langle$001$\rangle$ out-of-plane orientation, identified in advance with EBSD. The energy of the monochromatic beam was chosen to match that of a \{00n\} reflection in the range of 11 to 16 keV. This ensured that the dominant source of the diffraction signal is from the top 5 \textmu m of the sample where the irradiated layer lies. For each measurement, an energy range of $\sim$40 eV was scanned to fully capture the diffraction peak in 3D reciprocal space. An energy step size of 1 eV was used for all samples except for the lowest dose irradiation, where a 0.5 eV step size was used. It should be noted that measuring only the reflection closest to the normal orientation of the grain gives the out-of-plane strain, not the full strain tensor. Previous DAXM measurements on ion-irradiated tungsten showed that the in-plane strains are small compared to the out-of-plane strain \cite{Das2018a, Hofmann2015}. This is due to the constraint imposed by the unimplanted material beneath the irradiated layer.
		
		Two of the data points were further studied in a similar method depth-resolving the diffraction signal using a coded aperture instead of a platinum wire. Specifically these measurements were performed for Fe8Cr 0.8E-4 dpa and Fe8Cr 3.6 dpa samples. Further details of the coded aperture technique can be found elsewhere \cite{Gursoy2022, Gursoy2023}.
		
				
		\subsection{Nanoindentation}
		The hardness of the implanted layer was measured by nanoindentation using an MTS Nano Indenter XP with a diamond Berkovich tip (Synton-MDP). The tip area calibration was performed on fused silica (elastic modulus of 72 GPa). Indents of 2 \textmu m deep were performed using continuous stiffness measurement (CSM) mode \cite{Oliver1992}, with a strain rate of 0.05 s$^{-1}$, a CSM frequency of 45 Hz, and a harmonic amplitude of 2 nm.  At least 35 indents, spaced at a minimum distance of 50 \textmu m apart, were carried out across a minimum of 10 grains on each sample. 
		
		The representative hardness values were analysed in two ways. The first by taking the average hardness from CSM between indentation depths of 300--600 nm for all samples. This depth range was chosen to largely avoid indentation size effects (ISE) at shallower depths and contribution from the softer unirradiated bulk at greater depths. 
		
		The second method considers the Nix-Gao model, which accounts for the contribution of geometrically necessary dislocations to correct for ISE \cite{Nix1998}. The depth-dependence of hardness is described by:
		\begin{equation}
			\frac{H}{H_{0}} = \sqrt{1+\frac{h^{*}}{h}}
		\end{equation}
		where $H$ is the measured hardness at indentation depth $h$, $H_{0}$ is the bulk equivalent hardness at infinite depth, and $h^{*}$ is a characteristic ISE length scale that depends on the indenter shape and material hardness. This method has been previously applied to ion-irradiated ferritic alloys by limiting the indentation depth range over which the analysis is carried out to avoid contributions from the plastic zone extended into the unirradiated bulk \cite{Kasada2011, Zhu2024, Yang2024, Das2022}. For this study, the range of indentation depth for analysis is 100--600 nm. This minimises the effect of tip blunting and surface roughness at shallow indentation depths, and avoids the bulk unirradiated material contribution. We note that for the unirradiated sample, the indentation depth range for analysis is 100--2000 nm, making use of the full depth range.

		\section{Results}
		
		\subsection{Irradiation-induced lattice strain}
		
		\subsubsection{Extracting strain from diffraction data} \label{sec:extractstrain}
		\begin{figure}[h!]
			\centering
			\includegraphics[width=\textwidth]{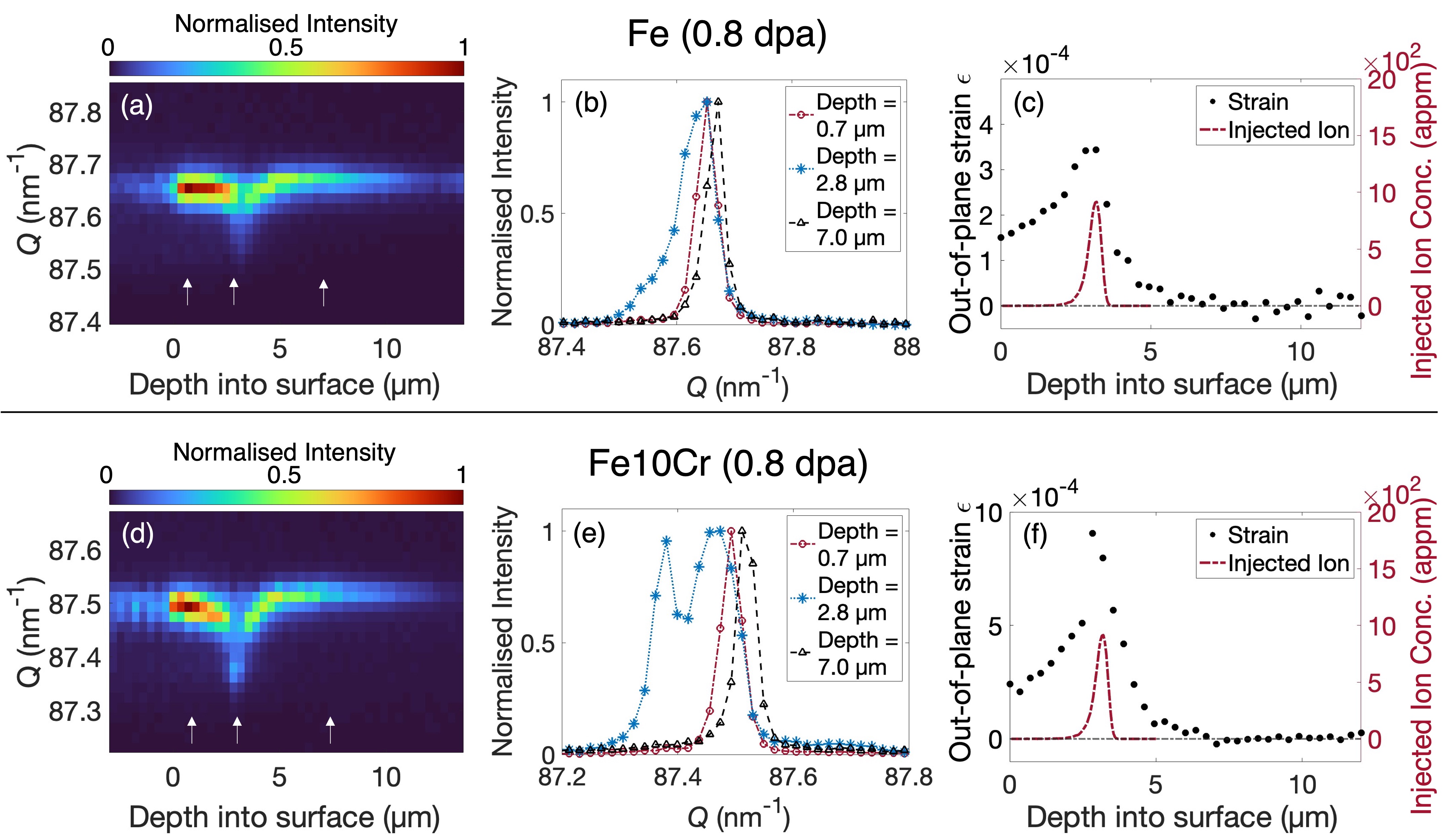}
			\caption{The integrated intensity plotted as a function of the magnitude of the scattering vector, $Q$, and depth perpendicular to the surface of the sample for (a) Fe at 0.8 dpa and (d) Fe10Cr at 0.8 dpa. (b), (e) Intensity vs. $Q$ for different depths into the sample (indicated by white arrows in (a) and (d)). The measured diffraction intensities shown here have been normalised to the point of highest intensity in each respective dataset. (c), (f) The corresponding lattice strain in the sample as a function of depth into the surface. The dashed line showing zero strain is included for comparison. }
			\label{fig:rawq}
		\end{figure}
		
		Figure \ref{fig:rawq}(a) and (d) show the measured integrated intensity for the (004) peak at 0.8 dpa in the Fe and Fe10Cr samples, respectively. The intensity is plotted as a function of the scattering vector magnitude $Q = |\boldsymbol{Q}|$ and of the reconstructed depth into the surface. The depth-reconstruction and integration procedure was performed with LaueGo \cite{Tischler2020a}. The distribution of $Q$ changes as a function of depth (Figure \ref{fig:rawq}(b) and (e)), with the distribution in the first 6 \textmu m shifting towards lower $Q$ values compared to those measured from 7 \textmu m and deeper. From Figure \ref{fig:dpa}, the predicted range of displacement damage and injected ions only extend to 3.5 \textmu m below the surface. Therefore, the distribution of $Q$ at depths much greater than this can be attributed to the undamaged material in each sample, conveniently serving as a built-in strain-free reference. 
		
		For each depth, a single Gaussian function was fitted to the intensity distribution in $Q$, and the centre of the distribution ($Q_{c}$) was extracted. An intensity-weighted average of $Q_{c}$ from depths 7 to 12 \textmu m was used as the strain-free reference $Q_{0}$ of each sample. The lattice strain $\epsilon$ at each depth is calculated using the expression:
		
		\begin{equation}
			\epsilon = \frac{Q_{0} - Q_{c}}{Q_{c}}.
		\end{equation}
		
		Performing this analysis for all depths yields the results shown in Figure \ref{fig:rawq}(c) and (f), respectively, for Fe and Fe10Cr implanted to 0.8 dpa. These plots show that irradiation causes lattice strain changes down to 6 \textmu m below the surface, which is significantly deeper than the damage layer thickness predicted by SRIM (Figure \ref{fig:dpa}). The strain profile increases with depth, reaching a peak at 3--3.5 \textmu m before decreasing to zero at a depths greater than 6 \textmu m. 
		
		Interestingly, at depths corresponding to the peak injected ion concentration, the intensity vs. $Q$ profile is not a single Gaussian distribution (depth  = 2.8 \textmu m--3.2 \textmu m in Figure \ref{fig:rawq}(b) and (e)). An additional peak appears at lower $Q$, corresponding to lattice expansion. This is presumably a result of injected ions existing as interstitials in the material, which have a positive relaxation volume, leading to lattice swelling \cite{Ma2019}. The appearance of an additional $Q$ peak was only observed for damage levels of 0.8 dpa or higher. Very clear splitting was observed for all FeCr alloys, while Fe samples only showed a slight asymmetric peak broadening.
		
		This study focuses on the lattice strain associated with the collision damage cascade formed during irradiation. As such, the subsequent analysis will focus on the measurements in the depth range of 0 to 2 \textmu m, where the effect of the injected ions is negligible.

		\subsubsection{Strain as a function of dose}
		
		\begin{figure}[h!]
			\centering
			\includegraphics[width=\textwidth]{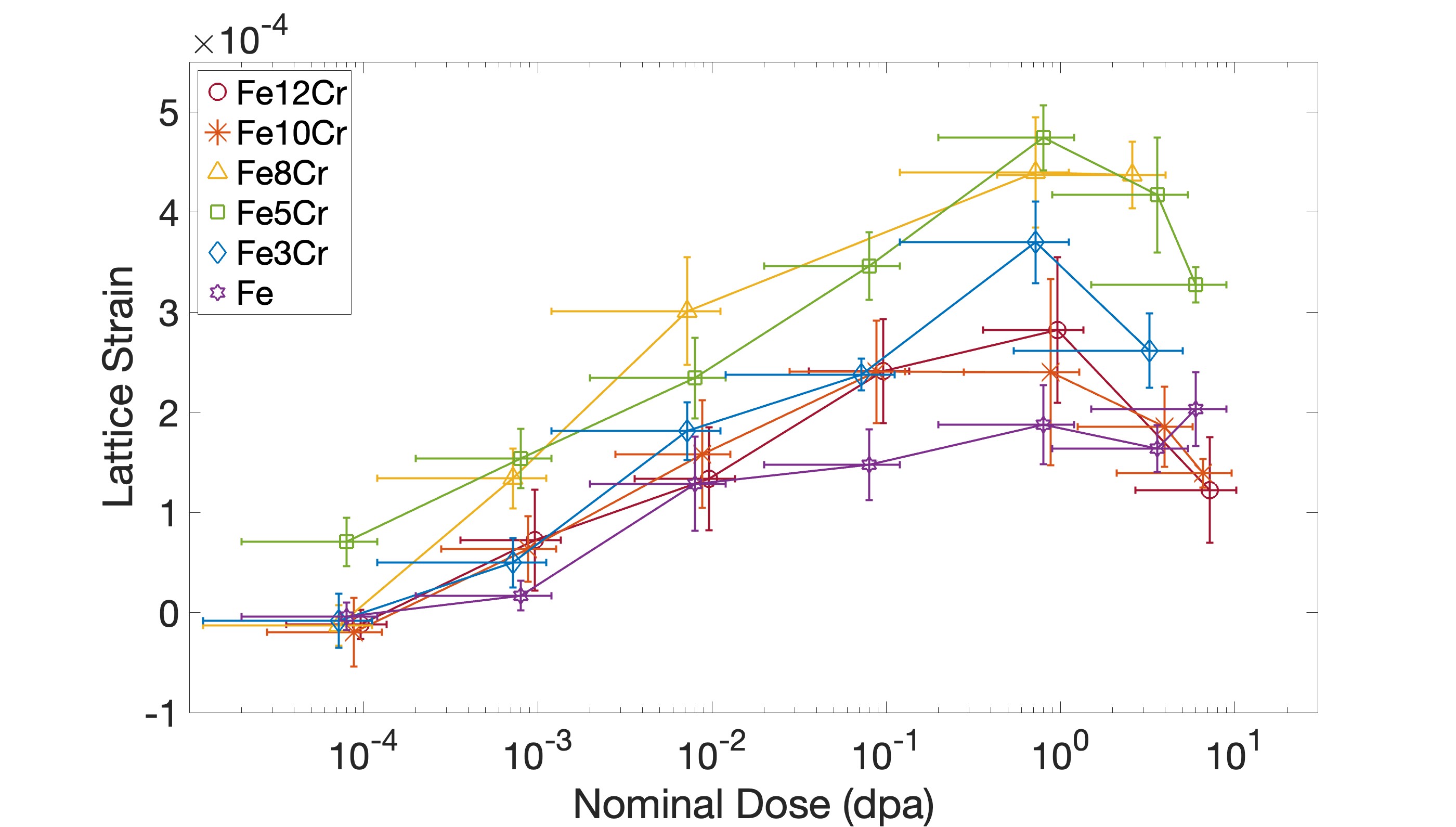}
			\caption{The average out-of-plane lattice strain in the top 2 \textmu m of each sample plotted as a function of irradiation dose for all compositions. The vertical error bars represent the standard deviation in lattice strain across the 2 \textmu m layer. The horizontal error bars represent the range of dose levels within the first 2 \textmu m of the samples. Markers have been offset horizontally for clarity.}
			\label{fig:totalstrain}
		\end{figure}
		
		The lattice strain caused by the displacement damage, as opposed to the injected ions, was examined by analysing the average strains in the top 2 \textmu m of the samples (Figure \ref{fig:totalstrain}). The reported uncertainties in lattice strains are the standard deviation values across the top 2 \textmu m of all measurements for each sample. The plotted error bars for dose represent the dose range in the top 2 \textmu m of the irradiated samples. For all sample compositions, except Fe5Cr, no lattice strain was detected at 0.8E-4 dpa. Lattice strain increases monotonically with dose from 0.8E-4 dpa to 0.8 dpa for all sample compositions. In the case of pure Fe, the strain does not change significantly as a function of dose beyond 0.8 dpa. However, for all FeCr alloys, there is a reduction in strain for doses greater than 0.8 dpa.
		
		Positive lattice strain indicates lattice expansion associated with crystal defects that have a positive relaxation volume ($\Omega$, in units of atomic volume). In BCC Fe, self-interstitials have a positive relaxation volume ($\Omega_{int} \sim$1.6--1.8), while vacancies have a negative relaxation volume ($\Omega_{vac}$ = -0.220) \cite{Ma2019}. Therefore, the net effect of a Frenkel pair is a positive relaxation volume. A reduction in lattice strain thus suggests that there is a removal of interstitial defects from the system with increasing irradiation dose, or a greater retention of vacancies than interstitials. 
		
		
		\subsubsection{Strain as a function of Cr}
		\begin{figure}[h!]
			\centering
			\includegraphics[width=\textwidth]{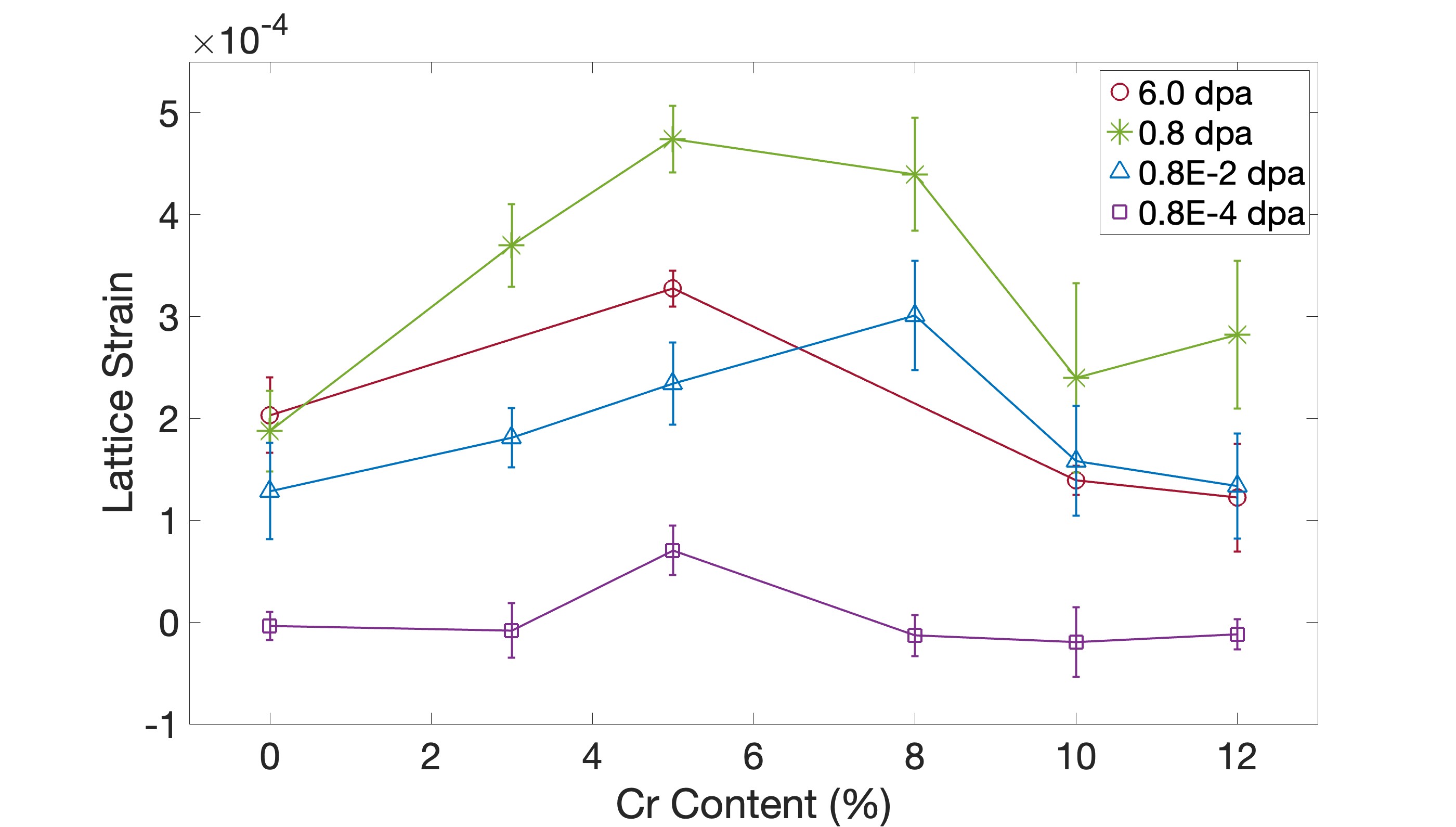}
			\caption{The average out-of-plane lattice strain in the top 2 \textmu m of each sample as a function of Cr content for 0.8E-4 dpa, 0.8E-2 dpa, 0.8 dpa and 6.0 dpa. The vertical error bars represent the standard deviation in lattice strain in the 2 \textmu m layer for all measurements taken for each particular composition and dose.}
			\label{fig:totalstrainCr}
		\end{figure}
		
		For all irradiation dose levels, a non-monotonic relationship between lattice strain and Cr content is observed (Figure \ref{fig:totalstrainCr}). Pure Fe samples exhibit the lowest lattice strain level compared to FeCr samples at any given dose, except 6.0 dpa. Lattice strain appears to be greatest for Fe5Cr and Fe8Cr at all measured doses. Fe10Cr and Fe12Cr generally show a similar amount of lattice strain at a given dose level. 
		

		\subsubsection{Onset of strain with dose}
		\begin{figure}[h!]
			\centering
			\includegraphics[width=0.6\textwidth]{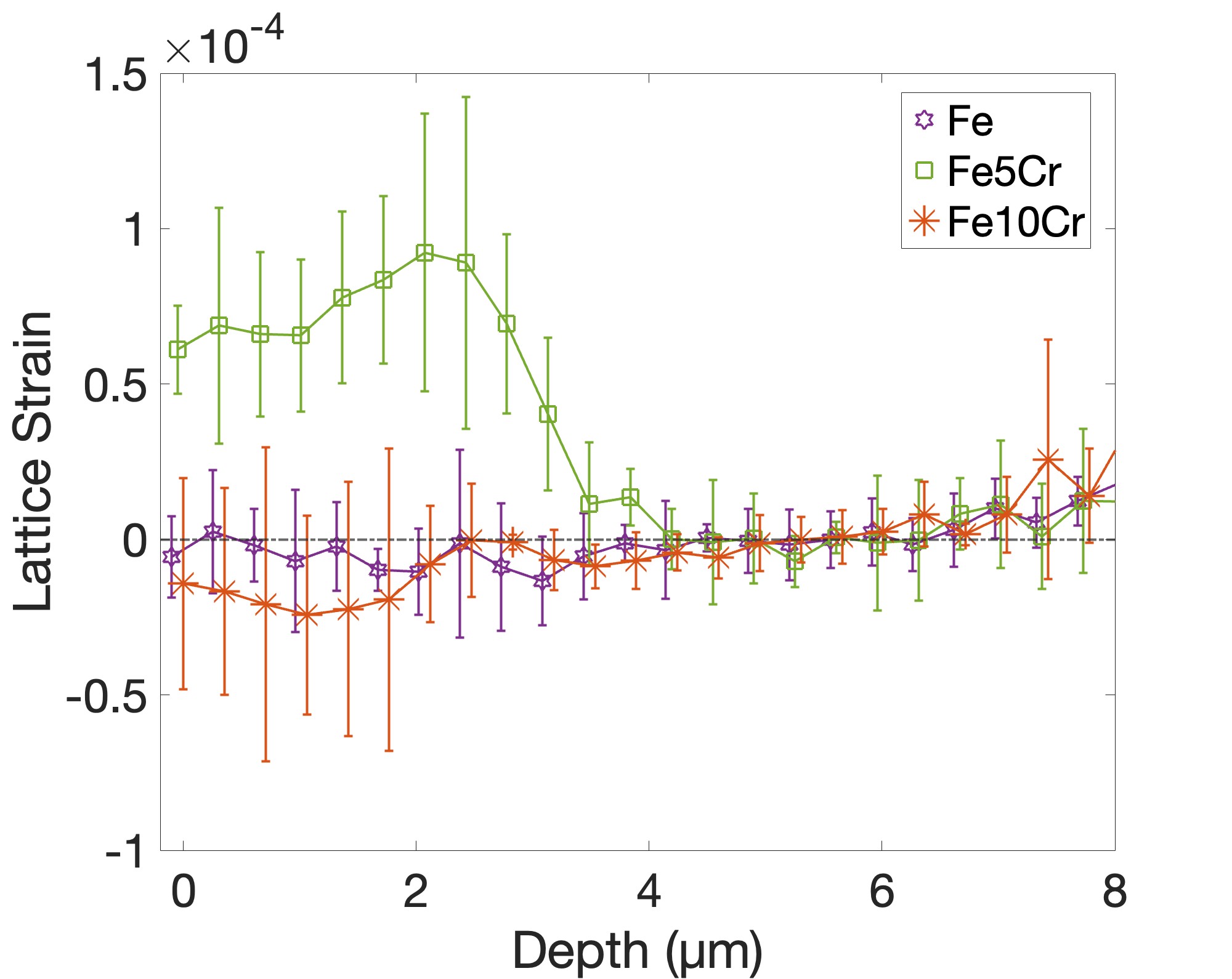}
			\caption{A comparison of the strain profile as a function of depth for Fe (purple), Fe5Cr (green), and Fe10Cr (orange) samples irradiated to 0.00008 dpa. The error bars represent $\pm$ 1 standard deviation from the measurements taken on each sample. Markers have been offset horizontally for clarity. A dashed line at zero strain is shown for reference.}
			\label{fig:lowestdose}
		\end{figure}
		
		At the lowest dose investigated for this study (0.8E-4 dpa), all samples except Fe5Cr showed no statistically significant strain within any depth of the implanted layer (some representative examples are shown in Figure \ref{fig:lowestdose}). There are some measurement points for Fe and Fe10Cr which appear to show negative strain. However, considering the size of the error bars, which represent $\pm$ 1 standard deviation from values across all measurements on each particular sample, these negative values are likely due to experimental uncertainty rather than an implantation effect. 
		
		Even though there are some fluctuations in the measurements and fitting of intensity vs. $Q$ at these low dose levels, the Fe5Cr sample clearly exhibits positive strain in the first 4 \textmu m (Figure \ref{fig:lowestdose}). The strain profile peaks between depths of 2.5 \textmu m to 3 \textmu m, similar to profiles of samples implanted to higher doses (Figure \ref{fig:rawq}(c) and (f)). This peak is likely due to the presence of injected Fe ions. From the measured strain level, a lower-bound estimate of the equivalent Frenkel pair density can be calculated (Appendix B). For the Fe5Cr sample at 0.8E-4 dpa, this value is $5.9\times 10^{24}$ m$^{-3}$ within the first 2 \textmu m below the surface. Note that this corresponds to a defect retention efficiency of 87\%, which is much higher than the expected 20--30\% (further discussed in the subsequent section). This high defect retention is likely due to a combination of low irradiation temperature reducing recombination rates and the high interstitial defect retention of Fe5Cr. The detailed discussions of the possible mechanism for the latter factor are provided in Section \ref{sec:creffects}.

		\subsubsection{Comparison to the NRT model}
		The defect concentration predicted by the NRT model \cite{Norgett1975}, which is used to calculate the dose in dpa, is a major overestimation, particularly at primary knock-on atom (PKA) energies much higher than the displacement energy of the material \cite{Nordlund2018}. This is the case for the present study where the irradiation was performed with 20 MeV ions, while the displacement energy of Fe is only 40 eV. The `NRT efficiency' is the ratio of the actual concentration of defects in the material, usually measured by electrical resistivity or calculated by molecular dynamics, to the NRT model predictions. From studies in the literature \cite{Nordlund2018}, the NRT efficiency of metals saturates at 0.2--0.3 at high PKA energy ($>$ 1 keV). 
		
		\begin{figure}[h!]
			\centering
			\includegraphics[width=\textwidth]{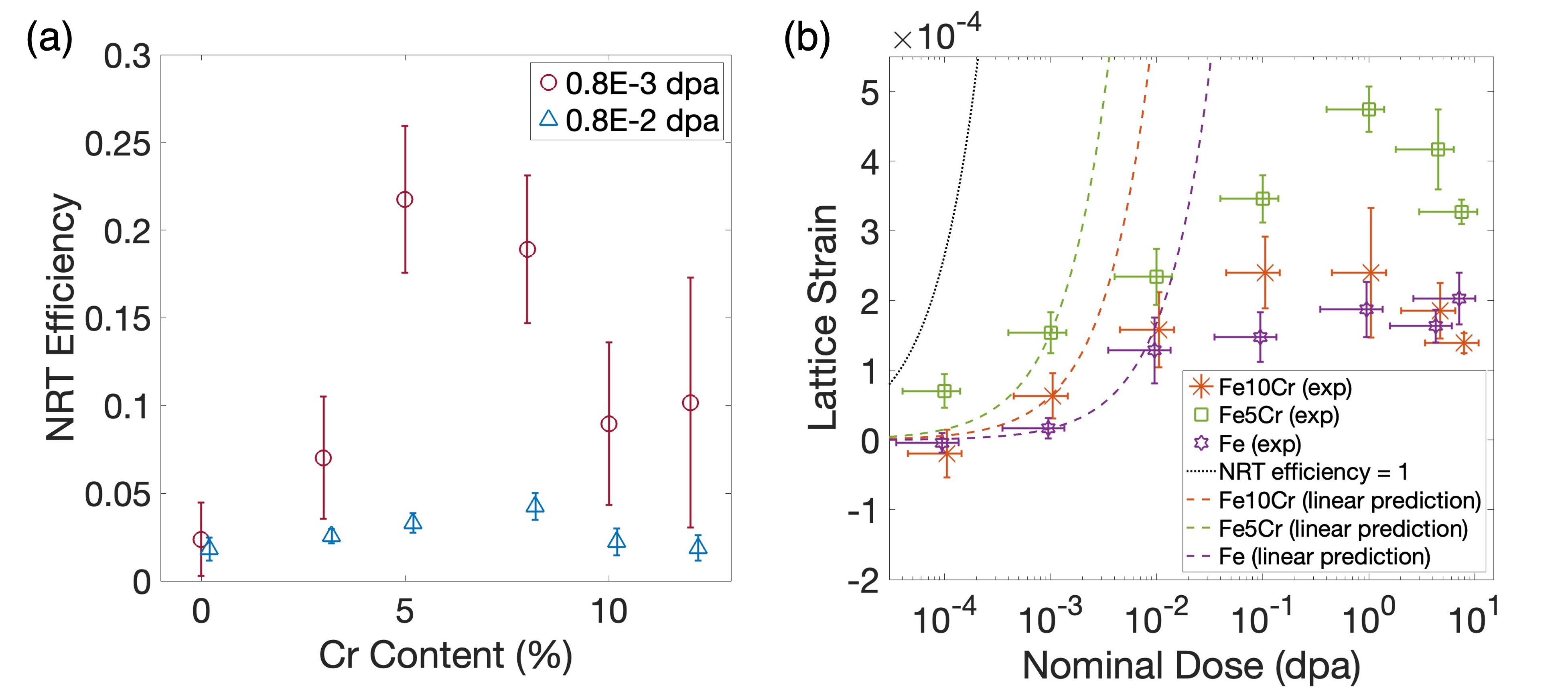}
			\caption{(a) The NRT efficiency as a function of Cr content for samples irradiated to 0.8E-3 dpa (red circles) and 0.8E-2 dpa (blue squares). (b) The measured lattice strain for Fe (purple), Fe5Cr (green) and Fe10Cr (orange) as a function of dose, compared to the predicted strain for a linear accumulation of isolated defects based on the NRT efficiency calculated at 0.8E-3 dpa (dashed lines of corresponding colours). The black dotted line shows the predicted strain profile in the case of 100\% defect production and retention (i.e. NRT efficiency = 1).}
			\label{fig:nrt}
		\end{figure}
		
		By converting the measured lattice strain into a corresponding Frenkel pair density (Appendix B), the NRT efficiency in Fe and FeCr at 313 K can be calculated (Figure \ref{fig:nrt}(a)). The lattice strain values at 0.8E-3 dpa and 0.8E-2 dpa were chosen for this calculation as all samples exhibit non-zero lattice strain at these doses. Uncertainties in the lattice strain measurements (from Figure \ref{fig:totalstrain}) have been propagated forward when calculating defect densities and NRT efficiencies. Low irradiation doses were chosen because the NRT-dpa definition and the NRT efficiency are only applicable for non-overlapping cascades \cite{Liu2021a}. Furthermore at low dose, the effect of defect clustering can be minimised, which allows a more accurate conversion of strain to defect density. This is important as one of the key assumptions made in the defect density calculation is that all defects present in the samples are isolated Frenkel pairs (Appendix B). 
		
		The NRT efficiency is highest for Fe5Cr and Fe8Cr at both doses examined (Figure \ref{fig:nrt}(a)), and the values fall within the expected range of 0.2--0.3 reported in the literature \cite{Nordlund2018}. It is the lowest for pure Fe at $\sim$0.02. 
		
		There is a reduction in NRT efficiency for all FeCr samples from 0.8E-3 dpa to 0.8E-2 dpa. This suggests either a decrease in defect retention, or the occurrence of significant defect clustering or cascade overlap. Interestingly, this is not the case for pure Fe, which has the same NRT efficiency at both doses within measurement error. 
		
		By comparing to the theoretical strain from 100\% defect production and retention (i.e. NRT efficiency = 1), it can be seen that our samples are in the regime where defect retention is much lower than predicted by the NRT model (Figure \ref{fig:nrt}(b)). For all FeCr samples, defect accumulation deviates from linear behaviour from 0.8E-3 dpa onwards. For Fe, this deviation happens beyond 0.8E-2 dpa. This suggests the `low dose' regime, before the onset of cascade overlap, ends between 0.001 to 0.01 dpa. This transition dose also depends on NRT efficiency, with a lower dose threshold for higher NRT efficiency material (such as Fe5Cr and Fe8Cr).


		
		
		\subsection{Hardness}
		
		\begin{figure}[h!]
			\centering
			\includegraphics[width=\textwidth]{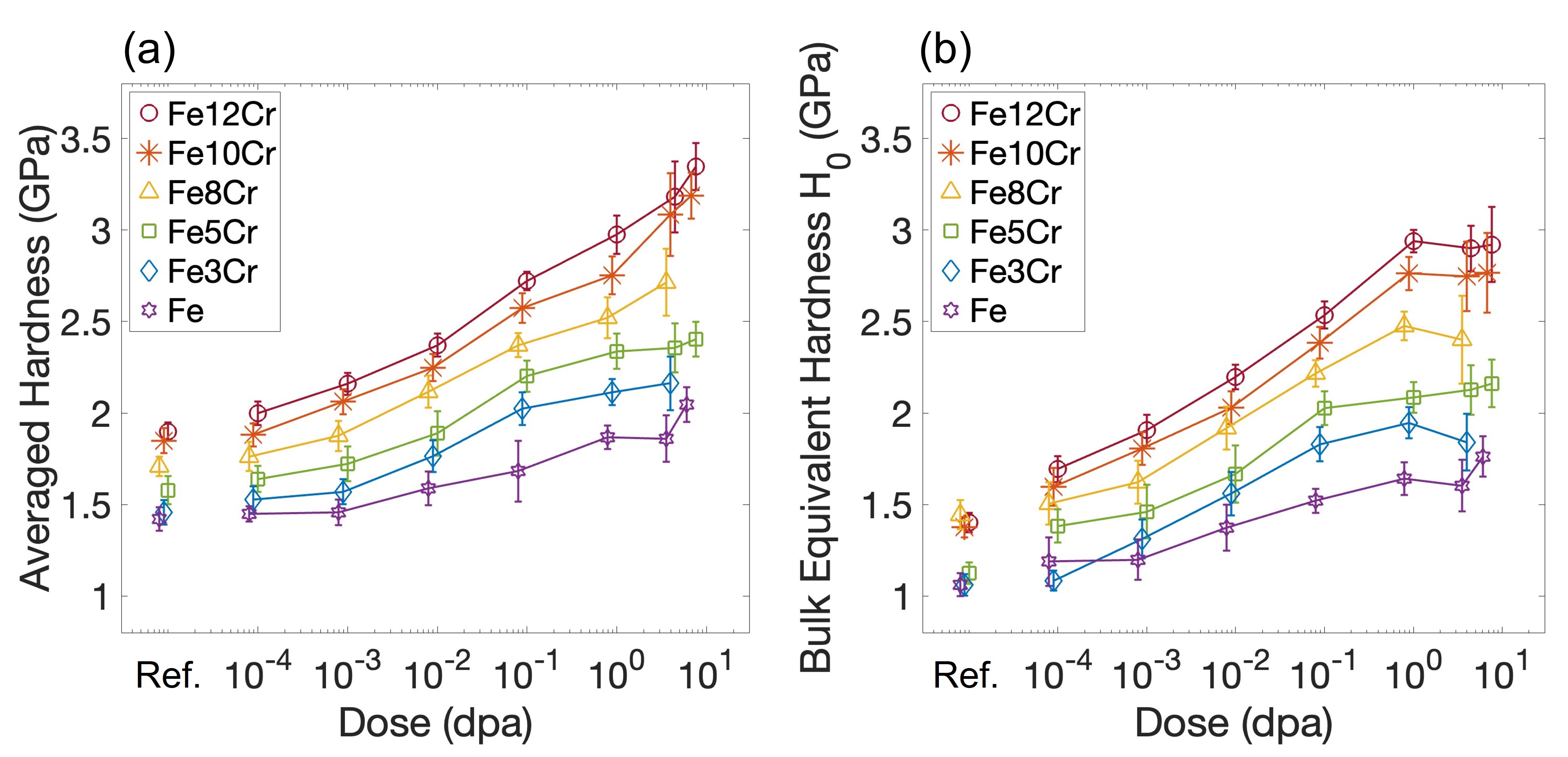}
			\caption{Hardness of samples analysed by (a) averaging between 300--600 nm (constant-depth analysis) and (b) using the Nix-Gao model, extracting the bulk equivalent hardness $H_{0}$. The unimplanted samples are included on the left side of each plot as a reference. The error bars represent $\pm 1$ standard deviation of all measurements performed on each sample. Markers have been offset horizontally for clarity.}
			\label{fig:hardness}
		\end{figure}
		
		Some solid solution hardening with the increase of Cr concentration is evident. From the constant-depth analysis (Figure \ref{fig:hardness}(a)), hardness appears to increase linearly with Cr content. However, from the Nix-Gao model (Figure \ref{fig:hardness}(b)), the bulk equivalent hardness shows that there appears to be negligible hardness change with the addition of up to 5\% Cr, which agrees with previous findings from the same bulk materials \cite{Song2020}. There also appears to be no change in hardness from 8\% to 12\% Cr. Overall, the bulk equivalent hardness $H_{0}$ is lower than the constant-depth hardness, demonstrating that there is still contribution to hardening due to ISE even when averaging between 300--600 nm indentation depth data. Data of the fitted ISE length scale ($h^{*}$) is included in Appendix C.
		
		
		The irradiation hardening increases as a function of dose up to 0.8 dpa. The difference in hardness between samples of different Cr content at each dose increases with dose. This corresponds to a greater irradiation hardening rate for samples of higher Cr content. 
		
		Beyond 0.8 dpa, the hardness curves appear to flatten out slightly for Fe, Fe3Cr, and Fe5Cr, suggesting the onset of saturation in irradiation hardening. However, for Fe8Cr, Fe10Cr, and Fe12Cr, the dose-dependence of hardening differs between the two analysis methods. The constant-depth analysis suggests an increase of hardness as a function of dose, whereas the Nix-Gao method suggests hardness saturation, similar to the samples of other compositions in this study.

		

		\section{Discussion}
		
		\subsection{Trends with dose}
		
		\subsubsection{Low dose effects ($<$ 0.008 dpa)} \label{sec:lowdose}
		The lattice strain for Fe5Cr observed at 0.8E-4 dpa (the lowest dose in this study) corresponds to a lower bound equivalent Frenkel pair density of $5.9\times 10^{24}$ m$^{-3}$. The dose threshold for defect observation with TEM in irradiated FeCr (manufactured under the same project as the samples in this study) has been reported as 1.5E-3 dpa  \cite{Schaublin2017}. This threshold is 20 times higher than the dose at which we observe the onset of defect-induced lattice strain. The authors of the TEM study estimated the defect density at 1.5E-3 dpa to be $6.5\times 10^{20}$ m$^{-3}$. From the lattice strain measurements at a similar dose in this study (0.8E-3 dpa, Figure \ref{fig:totalstrain}), the estimated defect density is between $6.0 - 17.1 \times 10^{24}$ m$^{-3}$ for Fe and FeCr alloys. The significant difference in defect density estimates from X-ray lattice strain measurements compared to TEM measurements may be attributed to the limited sensitivity of TEM to very small defect features ($<$ 2 nm) \cite{Zhou2006}. At low doses, these small defects are expected to dominate the defect population \cite{Jenkins2009, Yi2015}. 
		
		Similar findings from a different study of Fe irradiated with neutrons at 325--345 K revealed that the defect cluster density estimated by TEM is over two orders of magnitude lower than that determined by positron annihilation spectroscopy \cite{Zinkle2006}. From that study, a nanocavity density of $2\times10^{23}$ m$^{-3}$ at 1E-4 dpa was measured, with most nanocavities less than 1 nm in diameter (below TEM resolution limit). Even though this nanocavity density value is different from the equivalent Frenkel pair density found at 0.8E-4 dpa in this study, the difference can be reconciled by considering that an 1 nm nanocavity could contain an equivalent of up to 10 vacancy point defects \cite{Eldrup2002}.
		
		Irradiation hardening was observed for all Fe and FeCr samples at 0.8E-4 dpa. The amount of hardening increases with Cr content, from 2\% hardening for pure Fe to 5\% hardening for Fe12Cr. In the literature, a 10\% increase in yield stress (related to changes in indentation hardness) was observed for pure Fe following room temperature neutron irradiation to 1.2E-4 dpa, comparable to the lowest dose explored in this study \cite{Zinkle2006}. However, the TEM characterisation on the same sample did not identify any visible defects. 
		
		A neutron dose of $\sim$0.7E-3 dpa in Fe, Fe4Cr and Fe9Cr at the same temperature as the current study (313 K) has been found to cause an increase in yield stress of 35--40\% \cite{Hammad1982}. In comparison, at 0.8E-3 dpa, an increase of 3--13\% (increasing with Cr content) in nanoindentation hardness was measured in our study. However, a direct comparison of nanoindentation hardness and yield stress from bulk tensile testing is not straightforward due to the difference in lengthscale and mechanisms \cite{Johnson1970, Song2022}. Nanoindentation hardness depends on both the availability of sources for the nucleation of dislocations, and their subsequent propagation. On the other hand, macroscopic yield stress from bulk tensile testing depends mainly on the mobility of pre-existing dislocations. Another reason for the discrepancy between our study and that of Hammad \textit{et al.} \cite{Hammad1982} could be due to the different levels of impurities present. The carbon content for all of our samples is less than 10 wppm ($<$ 30 appm), according to the manufacturer \cite{Coze2007, Fraczkiewicz2011}. This value is 10 times lower than that reported in \cite{Hammad1982}. Lower impurity levels could greatly reduce defect retention, leading to less irradiation hardening in our samples. 
		
		The calculation of NRT efficiency at 0.8E-3 dpa for Fe5Cr (0.22) and Fe8Cr (0.19) is in good agreement with literature values (0.2--0.3) \cite{Malerba2006, Nordlund2018}. An important consideration in this comparison concerns the clustering of defects. In this study, the calculation of defect density relies on the assumption that defects remain as isolated Frenkel pairs. When clustering occurs, the relaxation volume per defect decreases, requiring more Frenkel pairs to be present to produce the same amount of strain as a corresponding population of isolated point defects. Hence the estimates of defect density and NRT efficiency can be interpreted as a lower-bound estimate. Furthermore, measurements of NRT efficiency in most other studies \cite{Nordlund2018} were performed at 4 K, which reduces the rate of defect recombination. Higher temperatures, such as in our study, therefore result in a further decrease of NRT efficiency \cite{Sahi2018}. The low levels of impurities present in our materials may further limit defect retention \cite{Castin2019}, which could contribute to a lower experimentally measured value of NRT efficiency compared to the other studies. 
		
		In the low dose regime (0.8E-4 dpa to 0.8E-3 dpa), an increase in irradiation dose leads to a monotonic increase in both lattice strain and hardness. Defects likely evolve from isolated Frenkel pairs to some clustering, resulting in a reduction in NRT efficiency with increasing dose. Eventually, the threshold dose is reached for direct observation of defects in TEM ($\sim$1E-3 dpa \cite{Schaublin2017}). 

		\subsubsection{Intermediate dose effects (0.008 $\leq$ dose $<$ 0.8 dpa)}
		
		The hardness and lattice strain measurements of this present study are in good agreement with previous measurements conducted on Fe3Cr, Fe5Cr and Fe10Cr (same raw materials) in similar conditions to doses of 0.8E-2 and 0.8E-1 dpa \cite{Song2020}. The lattice strain measurements of this study agree with previous measurements (averaged within the top 2 \textmu m of the sample) to within 30\%. Discrepancies can be attributed to measurement uncertainties for small strains and slight variations between individual grains. The hardness measurements of this study agree with those from \cite{Song2020} to within 9\%. The difference in hardness values can be attributed to the grain orientation specificity \cite{Spitzig1970} of the previous study, which only considers grains with $\langle 100\rangle$ out-of-plane orientation. In both cases, the evolution of hardness is similar with a monotonic increase with both dose and Cr content.
		
		In the intermediate dose regime, the rate of lattice strain increase decreases with increasing dose. This indicates a deviation from a linear accumulation of isolated Frenkel pairs ($\Omega$ $\sim$1.4--1.6). 
		This could result from the clustering of defects or a change in the ratio of interstitial to vacancy defects. Several factors likely contribute to this. 
		
		The first is the effect of cascade overlap, which becomes important at doses greater than $\sim$1E-2 dpa, leading to the formation of dislocation loops \cite{Yao2008}. This causes a reduction in the relaxation volume, and thus lattice swelling contribution, per defect. 
		Another factor is the reduction in the survival rate of subsequently introduced defects due to pre-existing defects in the crystal \cite{Gao1996, Sand2018}. This leads to an overall lower rate of defect population growth.
		
		The effect of finite temperature is also important. Stages of defect recovery as a function of temperature have been identified from resistivity studies in Fe \cite{Takaki1983}. At 313 K, the active mechanisms include Frenkel pair recombination, as well as di-interstitial and vacancy migration. 
		Since interstitials have much greater mobility than vacancy defects, they are more likely to cluster and reduce their contribution to overall lattice strain per defect. Furthermore, the increase in concentration of defects will also lead to a greater rate of recombination \cite{Takaki1983}. The net effect is the reduction in the rate of lattice strain increase, which is observed for all samples.

		
		\subsubsection{High dose effects ($\geq$ 0.8 dpa)}
		The reduction of lattice strain with increasing irradiation dose above 0.8 dpa in FeCr suggests a net removal of interstitial defects. This has been observed previously in self-ion irradiated tungsten by Mason \textit{et al.} \cite{Mason2020}. In tungsten, lattice strain increased monotonically up to 0.032 dpa, before dropping to zero at 0.056 dpa and ultimately becoming negative at doses beyond 1 dpa. This experimental result was compared to simulation results from Frenkel pair creation and insertion using the creation-relaxation algorithm (CRA). The CRA involves randomly displacing an atom to a new position within the simulation cell and then minimising the global potential energy in the cell \cite{Derlet2020}. Agreement was obtained between experiments and simulation results regarding the dose at which the strain peaked and then changed signs. But the simulation results predicted a strain level that was 10 times greater in magnitude than experimental observations. 
		
		For iron, Derlet and Dudarev have previously simulated strain induced by irradiation using CRA up to 2.5 cdpa (canonical dpa) \cite{Derlet2020}. Note that the definition of cdpa in CRA simulations is the ratio of the number of Frenkel pairs inserted to the total number of atoms in the simulation. This is an analogous measure of defect production to dpa. From CRA in iron, an increase of strain is observed up to 0.07 cdpa, peaking at a strain of $5 \times 10^{-3}$ before decreasing to zero strain beyond 2 cdpa. Compared to the experimental results of this study, the magnitude of strain predicted is 10 times greater than those experimentally observed in our study. The theoretical prediction of the dose threshold for maximum lattice strain is also a factor of 10 lower than we observe experimentally. These observations are similar to the results found for tungsten \cite{Mason2020}. 
		
		The phenomenon of a positive peak in strain followed by a trend towards zero strain (and in the case of tungsten, ultimately negative strain) at high doses is attributed to the formation of large-scale defect microstructures from an accumulation of interstitial defects \cite{Mason2020}. As the density of interstitial defects increases with dose, the defects begin to coalesce. The growth of these extended defect structures eventually causes an evolution back towards a less defective crystal structure, which removes positive lattice strain \cite{Derlet2020}. As vacancies are much less mobile, they largely remain as isolated defects, contributing negative lattice strain, which becomes dominant at high dose. 
		
		However, unlike tungsten, the lattice strain in iron and iron-chromium alloys does not become negative, either in simulations \cite{Derlet2020} or experimentally in this study. The reason for this could be the ratio of mobilities for interstitial and vacancy defects. In BCC iron, the migration energy of a single vacancy is 0.65 eV \cite{Domain2001} and for a self-interstitial loop, it can be as low as 0.1 eV \cite{Johnson1964}. Considering an Arrhenius rate behaviour for thermally-induced defect movement in crystals, there is a factor of $\sim$1.7 difference between interstitial and vacancy mobility. This is much lower than the corresponding ratio for tungsten ($>$ 5) \cite{Amino2016, Nguyen-Manh2006}. This means for iron, a higher rate of Frenkel pair recombination is expected. Furthermore, the rates of migration and annihilation of defects to sinks (e.g. dislocations and grain boundaries), and defect clustering would be more comparable between interstitial-type and vacancy-type defects. Therefore, even after the formation of extended dislocation networks at high doses, the imbalance of interstitial-type and vacancy-type defects is not sufficient to cause the net lattice strain in iron to become negative. However for tungsten, the low mobility of vacancies causes them to remain `frozen' in the lattice, such that they ultimately cause a negative net strain at high doses after the interstitial defects aggregate to form new crystal lattice planes \cite{Mason2020}.
		
		The formation of large extended defect microstructures in iron following room temperature high-dose ($>$ 6.5 dpa) irradiation has been previously observed by TEM \cite{Hernandez-Mayoral2008}. This dose is different to the dose threshold of strain reduction that resulted from defect coalescence in the present study. However, it's worth noting that TEM irradiation suffers from defect loss to the surface of the foil, which would delay the onset of defect coalescence \cite{Liu2021}. 
		
		
		The lattice strain predicted by CRA simulations of Fe \cite{Derlet2020} peaks at a dose 10 times smaller than the experimental observations in FeCr. This could be attributed to the greater mobility of defects in Fe and FeCr at 313 K which would lead to increased recombination and sink-annihilation of Frenkel pairs, thus delaying defect evolution and formation of extended structures. This temperature effect could also be the reason for the factor of 10 difference in the magnitude of strain between simulations and experiments. Since CRA simulations are athermal, with purely stress-driven relaxation of the lattice, there will be much higher levels of defect retention due to lower recombination rates. This is evident when comparing defect content and NRT efficiency between this experiments and CRA simulation. From CRA simulations, the ratio between defect content and cdpa value (analogous to NRT efficiency) for cdpa $<$ 0.01 is close to 1. Whereas from experiments, the NRT efficiency beyond 0.008 dpa is less than 0.1 for all compositions in this study (Figure \ref{fig:nrt}), in part due to temperature-enhanced recombination.
		
		It is worth noting that the presence of impurities in the experimental samples, even at low concentrations, compared to a perfect starting crystal in CRA simulations would cause enhanced retention of defects \cite{Terentyev2011, Niu2023, Hashimoto2014}. This effect would act in opposition to that caused by finite temperatures, as discussed previously. Interestingly, in our case, it appears that the temperature effect is dominant, resulting in a net delay in lattice strain evolution as a function of dose. This is consistent with the low impurity levels of the as-received materials in this study.
		
		The evolution of nanoindentation hardness is markedly different to the behaviour of lattice strain in this dose range above 0.8 dpa. We do not observe any changes in the sign of hardness change. Surprisingly, there do not appear to be any prior studies on hardness changes in FeCr as a function of dose following room temperature irradiation with self-ions. A study at room temperature following irradiation with Ar ions for RAFM T91 steel suggests hardness saturation at 4 dpa \cite{Karpov2019}. A study of CLAM steel irradiated with Xe ions at room temperature up to 5 dpa showed no hardness saturation \cite{Chang2014}. However, direct comparison with the results of the present study is not straightforward due to the use of noble gases as irradiating particles in these previous works. 
		
		Self-ion studies \cite{Heintze2011, Hardie2013} have only been performed for irradiation at 573 K and showed hardness saturation for pure Fe and FeCr alloys with Cr content greater than 5\% above 1--2 dpa. For low Cr content, no saturation was observed even up to 10 dpa. The trends from these studies conducted at higher irradiation temperatures differ from our results. Following room temperature irradiation, Fe, Fe3Cr and Fe5Cr approached hardness saturation at doses $\geq$ 0.8 dpa, but conclusive trends cannot yet be determined for Fe8Cr, Fe10Cr, and Fe12Cr. One reason for the discrepancy could be due to some carbon contamination from the irradiation process. Atom probe tomography data of Fe10Cr at 0.8 dpa is included in Appendix D. We estimate the carbon enrichment to be $\sim$100 appm (20 wppm) after 0.8 dpa, compared to 18 appm (4 wppm) reported by the manufacturer for the as-made samples. The additional carbon introduced, which scales with irradiation dose, could contribute in part to the hardening seen at high doses \cite{Wang2017}, particularly when analysed with the constant-depth method. 
		
		Another reason for the observed difference in hardening trends with other studies at higher irradiation temperatures could be enhanced defect mobility, which leads to more defect recombination. TEM studies have shown the movement of dislocation loops at 573 K around 0.6 dpa \cite{Hernandez-Mayoral2008}. Even at lower doses, the same authors observed a much greater fraction of mobile loops at 573 K compared to at room temperature \cite{Yao2008}. As such, one might expect the asymptotic or saturation state of defects in irradiated Fe and FeCr to occur at a lower dose for irradiation at higher temperatures.
				
		In contrast to lattice strain trends in the high dose regime, the non-negative change of hardness as a function of irradiation dose indicates that the presence of all irradiation defects, both isolated and within aggregated structures such as dislocation loops and networks, contributes to the hardening of Fe and FeCr. 

		\subsection{Trends with Cr content}\label{sec:creffects}
		The trends for lattice strain and hardness as a function of Cr content differ greatly. While an increase in Cr content at each dose is associated with a greater increase in hardness, lattice strains peak between Fe5Cr and Fe8Cr at each dose, with lower strains for higher and lower Cr content. As previously discussed, hardness is an indicator of the overall defect population, with interstitials and vacancies additively contributing. Lattice strain is dependent on the imbalance between interstitial-type and vacancy-type defects as their respective contributions (relaxation volume) have opposite signs. By examining both trends, we can gain an insight into how the defect population is affected by Cr content.
		
		There is a monotonic increase in hardness and hardening rate of Fe and FeCr with Cr content. This suggests that defect retention increases with Cr content, which causes an increase in defect number density and/or size \cite{Spatig2019}. Heintz \textit{et al.} \cite{Heintze2011} measured hardness after room temperature irradiation to 1 dpa and similarly found that hardening increased with Cr content. Interestingly, no hardening was observed for Fe2.5Cr in that study, which is different to our results.
		
		In contrast, the lattice strain trend is non-monotonic with maximum lattice strain at Cr content between 5--8\%. This suggests that the defect population and types depend on Cr content. Larger positive lattice strain can arise from a larger population and/or less clustering of interstitial-type defects, as well as a smaller population and/or more clustering of vacancy-type defects. Since interstitials are more mobile than vacancies in iron \cite{Domain2001}, their population and mobility may have a greater effect on lattice strain at room temperature. 
		
		It is interesting to note that FeCr has many non-monotonic trends in the range of Cr content less than 20\%. One such trend is that the change in ductile-to-brittle-transition temperature reaches a local minimum at 9\%Cr \cite{Kayano1988}. For irradiation-induced void swelling, it also varies non-monotonically with Cr content \cite{Bhattacharya2016, Little1979, Gelles1982, Garner2000, Lin2021}. The diffusivity of interstitial defects in FeCr is also a non-monotonic function of Cr content, as identified by molecular dynamics and Monte Carlo studies \cite{Malerba2013}. At low Cr concentrations, an increase in Cr content will increase the binding of self-interstitial clusters to the Cr atoms, reducing their mobility \cite{Arakawa2004}. With further increase of Cr content past a critical minimum point, each interstitial atom could be interacting with more than one Cr atom simultaneously, effectively `pulling' the defect in opposite directions. This then results in an increase of interstitial diffusivity with Cr content. The point of minimum interstitial diffusivity occurs around 10\%Cr \cite{Terentyev2005}. This mechanism has also been used to explain the local minimum of void swelling at 5\%--10\%Cr observed in some neutron irradiation studies (653 K, 30 dpa) \cite{Terentyev2005, Little1979}. It is proposed that the origin of void swelling lie in the absorption of fast-moving interstitial defects at sinks, such as grain boundaries and surfaces, leaving behind vacancies and vacancy clusters that subsequently form voids. A reduction in interstitial cluster mobility will limit their migration to sinks and thus enhance recombination, thereby reducing void swelling. Furthermore, Cr only weakly interacts with vacancies \cite{Olsson2007}, so their population is not affected as strongly by Cr content.
		
		Relating this to our observations, the balance of interstitial to vacancy defects will be strongly correlated with the mobility of interstitial defects. A reduction in the mobility of interstitial clusters will increase the ratio of interstitial to vacancies in the material, leading to a greater, positive lattice strain. The range of Cr content (5--8\%) that correspond to the highest level of lattice strain in this study is consistent with the range for minimum void swelling and cluster diffusivity \cite{Little1979, Malerba2013}, suggesting this could be an explanation for the non-monotonic dependence of lattice strain on Cr content. However, we note that trends for void swelling strongly depends on irradiation temperature \cite{Lin2021}, dose \cite{Bhattacharya2016, Gelles1982}, and sample processing history (e.g. cold-worked or annealed) \cite{Sencer2000}. Further systematic studies on void swelling could reveal more insights into the role of Cr on defect populations.
		
		For pure Fe, a delayed onset of lattice strain saturation is observed compared to the FeCr alloys. Even at 6.0 dpa, the lattice strain does not appear to have reached a peak, unlike for the other FeCr alloys where a decrease of lattice strain occurs beyond 0.8 dpa (Figure \ref{fig:totalstrain}). The NRT efficiency for all FeCr alloys is reduced between 0.8E-3 dpa to 0.8E-2 dpa (Figure \ref{fig:nrt} (a)). This suggests defect clustering is occurring, as it causes a reduction in strain contribution per defect, which leads to an underestimation of the defect density in the material. For pure Fe, the NRT efficiency remains constant between 0.8E-3 dpa to 0.8E-2 dpa, corresponding to a linear increase in isolated defect density. This suggests that pure Fe should exhibit defect clustering at a higher dose than FeCr. Furthermore, due to the low impurity content in the Fe material, the barrier to defect recombination is low, resulting in a low NRT efficiency and defect retention. As a result, all stages of defect evolution in Fe are shifted to higher doses compared to FeCr where defect retention is higher due to the binding of interstitial defects to Cr, as discussed earlier. 

		


		\section{Summary and Conclusion}
		In this study, the effect of ion irradiation at 313 K on lattice strain and hardness in Fe and FeCr alloys was investigated. A dose range of 0.00008 dpa to 6.0 dpa was covered for pure Fe and FeCr up to 12\%Cr. The key findings are as follows:
		\begin{itemize}
			\item Irradiation hardening was observed for all Fe and FeCr alloys at a dose as low as 0.00008 dpa. Non-zero lattice strain in the implanted layer was also measured at that dose for Fe5Cr, which corresponds to an equivalent Frenkel pair density of $5.9 \times 10^{24}$ m$^{-3}$. This is well below the dose limit for any defect detection reported in the literature by electron microscopy.
			\item The NRT efficiency was calculated for all alloys at 0.0008 dpa. The highest values ($\sim$0.2) were found for Fe5Cr and Fe8Cr, which agree with studies in the literature at similar ion energies. 
			\item FeCr alloys reach a maximum positive strain at a dose of 0.8 dpa. Further increase in dose caused a reduction in lattice strain due to the formation of extended interstitial defect structures. This agrees with simulation results after accounting for the effects of finite temperature and impurities. 
			\item A delay in the onset of interstitial clustering as a function of dose is observed in Fe compared to FeCr. No maximum in lattice strain was observed in Fe even at 6.0 dpa. In the absence of Cr, there is comparatively little defect pinning in Fe, especially given the high purity in the as-received material. As a result, the NRT efficiency is low. Combined with the temperature, this means the evolution of lattice strain in Fe at room temperature is delayed to significantly higher doses than in FeCr.
			\item Fe5Cr and Fe8Cr exhibit the highest level of positive strain out of all FeCr samples studied at any given dose level. This falls into the range of Cr concentration where the diffusivity of self-interstitial clusters is lowest, increasing their retention.
			\item There is a monotonic increase in hardness with irradiation for all samples investigated. However, even by 6.0 dpa, irradiation hardening has not reached saturation for Fe or FeCr alloys. Fe, Fe3Cr and Fe5Cr exhibit very low hardening rates while Fe8Cr, Fe10Cr and Fe12Cr exhibited higher hardening rates at doses beyond 0.08 dpa. 
		\end{itemize}

		\section*{Declarations}
		\subsection*{Funding}
		The  authors  acknowledge  use  of  characterisation  facilities  within  the  David  Cockayne  Centre for  Electron  Microscopy,  Department  of  Materials,  University  of  Oxford,  alongside  financial support  provided  by  the  Henry  Royce  Institute (grant renumber EP/R010145/1). This research used resources of the Advanced Photon Source, a U.S. Department of Energy (DOE) Office of Science User Facility operated for the DOE Office of Science by Argonne National Laboratory under Contract No. DE-AC02-06CH11357. This material is based upon work done at Brookhaven National Laboratory, supported by the U.S. Department of Energy (DOE), Ofﬁce of Basic Energy Sciences, under Contract No. DE SC0012704. KS acknowledges funding from the General Sir John Monash Foundation and the University of Oxford Department of Engineering Science. FH and DY acknowledge funding from the European Research Council (ERC) under the European Union’s Horizon 2020 research and innovation programme (grant agreement number 714697). DEJA acknowledges funding from EPSRC (grant number EP/P001645/1).
		
		\subsection*{Conflicts of interest}
		The authors have no relevant financial or non-financial interests to disclose.
		
		\subsection*{Data and code availability}
		All data, raw and processed, as well as the processing and plotting scripts are available at: \textit{A link will be provided after the review process and before publication.}
		
		\section*{Acknowledgements}
		The authors are grateful to Sergei Dudarev (United Kingdom Atomic Energy Authority) for his helpful discussions. The authors are also grateful to Andy Bateman and Simon Hills (Department of Engineering Science, University of Oxford) for their assistance with making the temperature-controlled sample holder for ion implantation.
		
		\renewcommand{\thefigure}{A-\arabic{figure}}
		\setcounter{figure}{0}
		\section*{Appendix A - EBSD maps of samples}
		Electron backscatter diffraction (EBSD) was carried on a Zeiss Merlin FEG-SEM at the David Cockayne Electron Microscopy Centre at the University of Oxford. The acceleration voltage used was 30 kV with a probe beam current of 10 nA. Post-processing was performed using the Oxford Instruments HKL Channel 5 Tango software to remove noise and determine the grain sizes.
		
\begin{figure}[h!]
			\centering
			\includegraphics[width=0.95\textwidth]{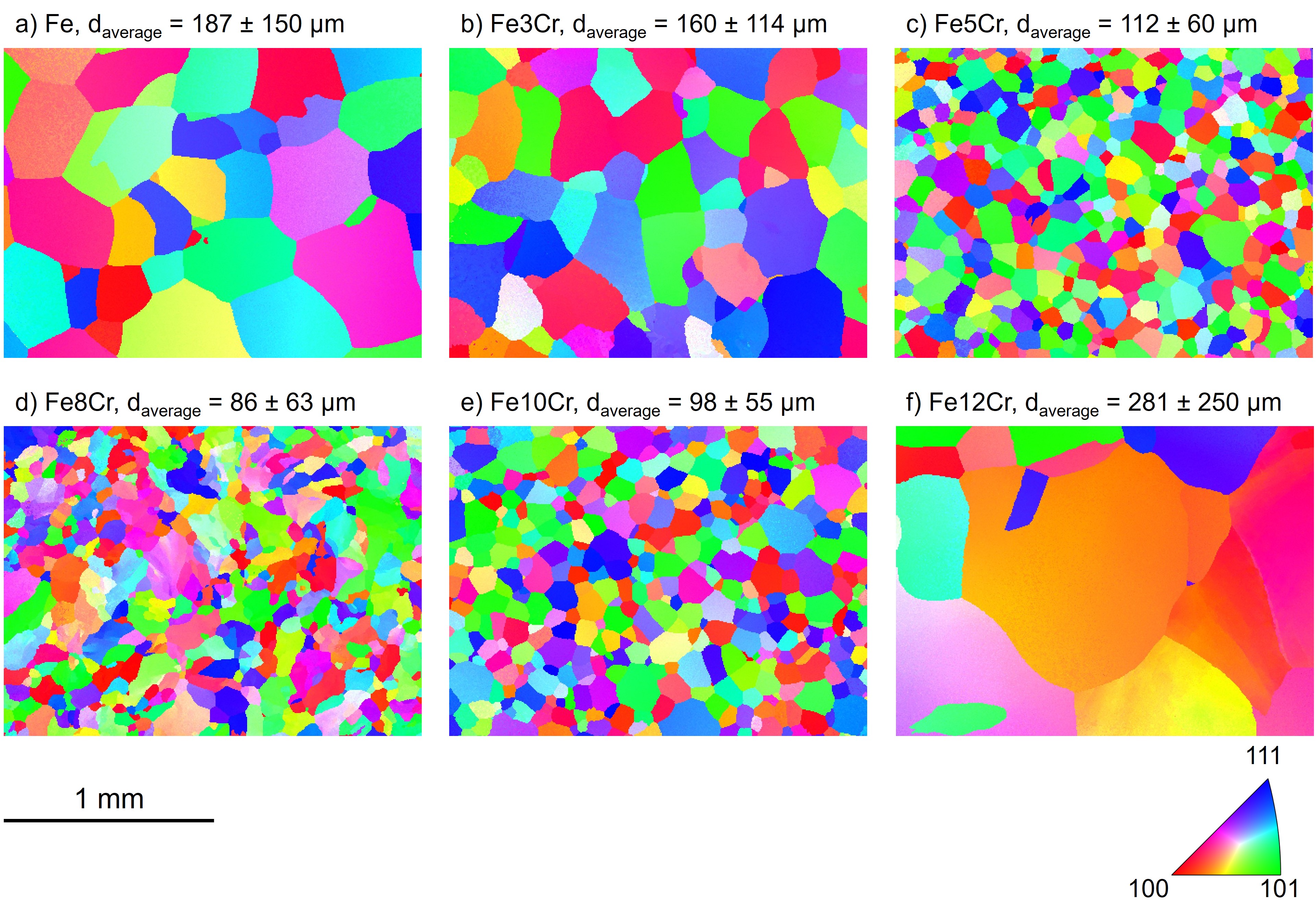}
			\caption{Representative grain maps for each composition of FeCr in this study measured by EBSD.}
			\label{fig:ebsd}
		\end{figure}

		The representative orientation maps (Figure \ref{fig:ebsd}) were used to select the appropriate areas for lattice strain measurements. The microstructure of the Fe8Cr material also exhibits signs of cold-work (intragranular misorientation). However, this did not have a significant impact on the lattice strain measurements.
		
		\newpage
		
		\renewcommand{\theequation}{B-\arabic{equation}}
		\setcounter{equation}{0}  
		\section*{Appendix B - Defect density calculations}
		The method used to calculate the defect density from the lattice strain measurements is similar to the method used in our previous study \cite{Song2020}. For further details and example calculations, readers are referred to the supplementary file of \cite{Song2020}. In this section, we provide an overview of the equations used and assumptions made in our defect calculations.
				
		\subsection*{B1 - Defect Density from Lattice Strain Measurements} 
		The out-of-plane strain, $\epsilon_{zz}$, induced by irradiation defects can be expressed as  \cite{Hofmann2015}:
		\begin{equation}\label{eqn:strain}
			\epsilon_{zz} = \frac{1}{3}\frac{(1+\nu)}{(1-\nu)} \sum_{A} n^{(A)} \Omega_{r}^{(A)}
		\end{equation}
		where $\nu = 0.3$ is the Poisson ratio (for pure Fe \cite{Cardarelli2018}), $n^{(A)}$ and $\Omega_{r}^{(A)}$ are respectively the number density and relative relaxation volume for each type of defect $(A)$. 
		
		For low irradiation dose ($\leq$ 0.008 dpa), we made the following assumptions in our calculations:
		\begin{itemize}
			\item There is no clustering of interstitials or vacancies. This means the relaxation volume per point defect is maximised \cite{Mason2019}.
			\item There is no loss of defects, particularly interstitials, to the surface of the materials or to sinks such as grain boundaries. This means that there is an equal number of interstitial atoms ($n^{i}$) and vacancies ($n^{v}$).
		\end{itemize}
		Using these assumptions, we obtain a lower bound of the equivalent Frenkel pair density ($n^{FP}$). Clustering of interstitials would decrease the relaxation volume per point defect \cite{Ma2019}, requiring more Frenkel pairs to be present in order to produce the same amount of positive strain that was measured (covered by assumption 1). If interstitials were lost to the surface, more equivalent Frenkel pairs would need to be present to account for the amount of positive strain measured (covered by assumption 2).
		
		Rearranging Equation \ref{eqn:strain} and multiplying by the atomic density of Fe ($\rho_{Fe} = 8.48 \times 10^{28}$ m$^{-3}$) yields the volumetric number density of equivalent Frenkel pairs:
		\begin{equation}\label{eqn:density}
			N^{FP} = \frac{\rho_{Fe}\epsilon_{zz}}{\Omega_{r}^{FP}} \left(\frac{3(1-\nu)}{(1+\nu)}\right)
		\end{equation}
		
		The relative relaxation volume, per point defect, of a $\langle 111 \rangle$ interstitial is $\Omega_{r}^{\langle 111 \rangle} = 1.65$ and that of a $\langle 100 \rangle$ interstitial defect is $\Omega_{r}^{\langle 100 \rangle} = 1.86$ \cite{Ma2019}. For a vacancy, the relative relaxation volume is $\Omega_{r}^{v} = -0.22$ \cite{Ma2019}. As a $\langle 100 \rangle$ interstitial has a higher positive relaxation volume, we can use $\Omega_{r}^{FP} = 1.86 - 0.22 = 1.64$ to get a lower bound estimate on the equivalent Frenkel pair density in the irradiated samples.
		
		\subsection*{B2 - NRT Efficiency Calculation}
		The NRT efficiency describes the ratio of produced and retain defect density to the value predicted by the NRT-dpa model \cite{Nordlund2018}. The calculation of the NRT-dpa was from the vacancy.txt file generated by SRIM using the Quick K-P model, following the descriptions provided in \cite{Stoller2013, Agarwal2021}. The NRT efficiency can then be expressed by:
		\begin{align}
			\nonumber \text{NRT efficiency} &= \frac{n^{FP}}{dpa} \\ 
			&= \frac{\epsilon_{zz}}{(dpa)\Omega_{r}^{FP}} \left(\frac{3(1-\nu)}{(1+\nu)}\right)
		\end{align}
		
		From Figure \ref{fig:nrt}, the NRT efficiency value for each sample was calculated from the average NRT-dpa in the top 2 \textmu m of each implantation dose profile. This is consistent with the calculation of the nominal dose for each sample.

		\renewcommand{\thefigure}{C-\arabic{figure}}
		\setcounter{figure}{0}
		\renewcommand{\thetable}{C-\arabic{table}}
		\setcounter{table}{0}
		\section*{Appendix C - Nix-Gao Analysis}
		
\begin{figure}[h!]
			\centering
			\includegraphics[width=0.9\textwidth]{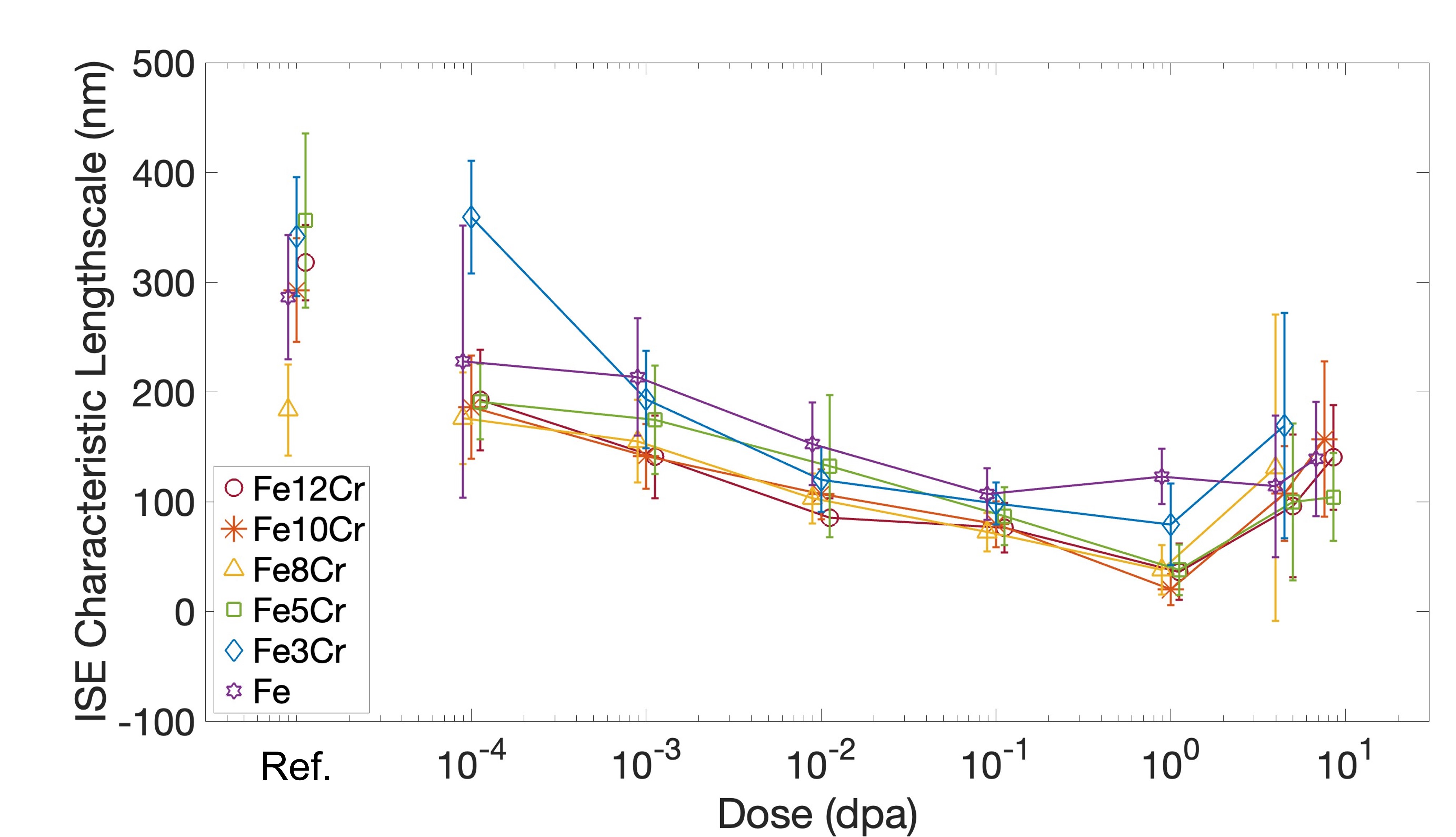}
			\caption{$h^{*}$ as a function of dose for different compositions.}
			\label{fig:hstar}
		\end{figure}

		The fitted $h^{*}$ parameter, representing the ISE characteristic length scale as a function of dose is shown in Figure \ref{fig:hstar}. $h^{*}$ appears to decrease with increasing dose and does not show a clear dependence on Cr content. We note that the fitting for 3.6 dpa and 6 dpa samples appear to show quite large errors compared to the lower dose samples.
		
		\newpage
		
		\renewcommand{\thefigure}{D-\arabic{figure}}
		\setcounter{figure}{0}
		\renewcommand{\thetable}{D-\arabic{table}}
		\setcounter{table}{0}
		\section*{Appendix D - Atom probe tomography results}
		Due to concerns about possible carbon contamination during the irradiation process, atom probe tomography (APT) was performed on the Fe10Cr sample, exposed to 0.8 dpa, to quantify the concentration of carbon present in the sample. APT analysis was performed using a Leap 5000XR microscope in laser mode. The laser energy was 50 pJ with a pulse frequency of 200 kHz and detection rate of 0.5. Samples were maintained at 50 K for the analysis.
		
\begin{figure}[h!]
			\centering
			\includegraphics[width=0.6\textwidth]{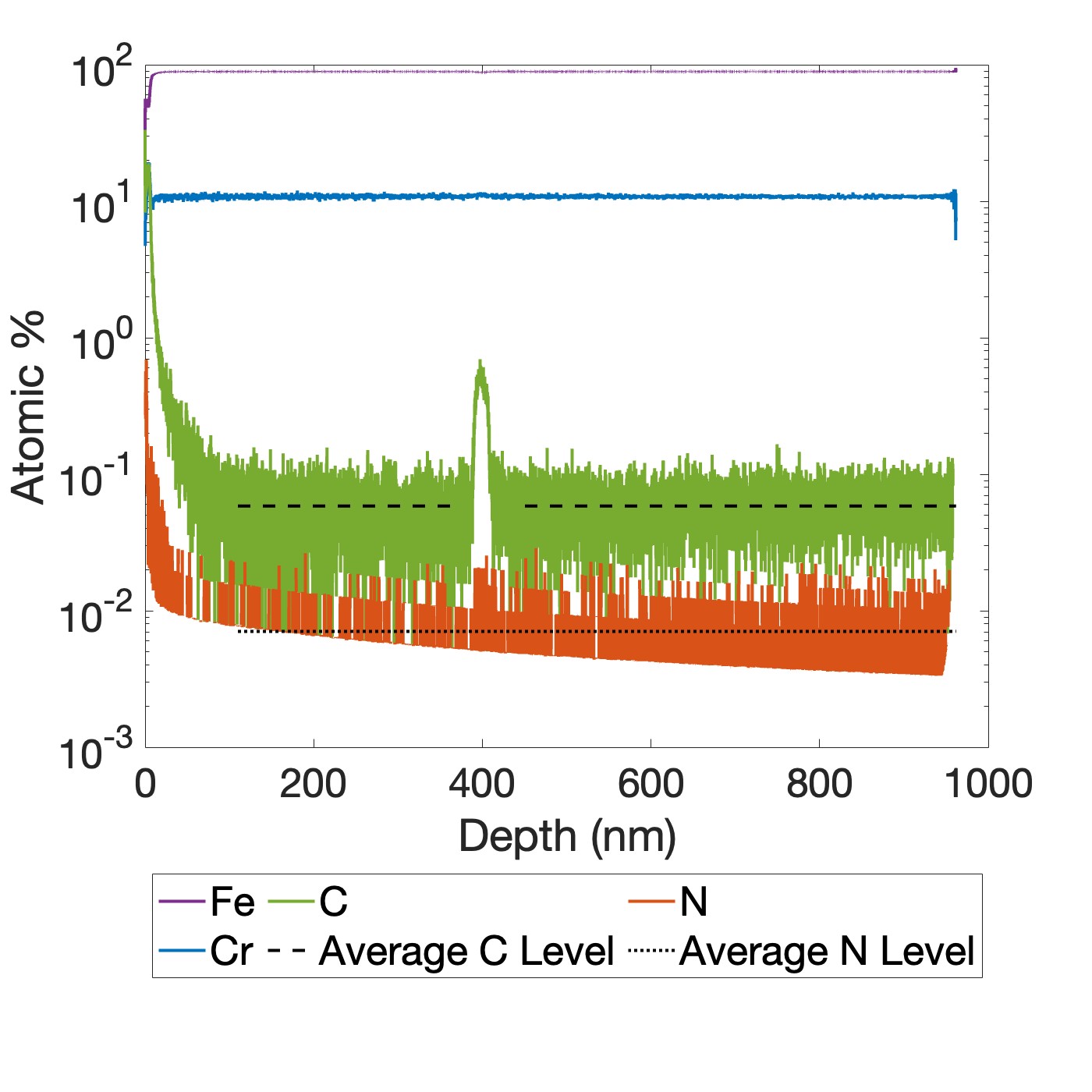}
			\caption{Depth profile of iron, chromium, carbon and nitrogen as measured by APT from the Fe10Cr 0.8 dpa sample.}
			\label{fig:APT}
		\end{figure}

		The depth profiles of a few key elements are presented in Figure \ref{fig:APT}. Near the surface (0 $<$ depth $<$ 100 nm), there is a high concentration of carbon and nitrogen, possibly due to surface contamination and FIB milling process. There is a spike in carbon concentration at $\sim$400 nm depth, which also correlated with a spike in oxygen atoms and C-Cr ions (not shown in plot). The averaged quantities are shown in Table \ref{tab:APT}.
		
	\begin{table}[h!]
		\begin{center}	
		\begin{tabularx}{\textwidth}{c *{4}{Y}}
			\toprule
			\multirow{2}{*}{Elements}
			& \multicolumn{2}{c}{Nominal content from manufacturer}  
			& \multicolumn{2}{c}{Measured content from APT}\\
			\cmidrule(lr){2-3} \cmidrule(l){4-5}
			& appm & wppm & appm & wppm \\
			\midrule
			C	& 18  & 4 & 585 $\pm$ 236 	& 126 $\pm$ 50	\\
			\addlinespace
			N	& 12  & 3 & 71 $\pm$ 37 		& 18 $\pm$ 9	\\
			\addlinespace
			O	& 14  & 4 & 160 $\pm$ 107 	& 46 $\pm$ 31	\\
			\addlinespace
			P	& $<$ 9 & $<$ 5 & 103 $\pm$ 84 & 58 $\pm$ 47	\\
			\bottomrule
		\end{tabularx}
		\caption{A comparison of the concentration (in both atomic and weight content) of different elements in the Fe10Cr alloy as reported by the manufacturer and measured by APT (averaged between 100 nm to 950 nm below the surface).}
		\label{tab:APT}
		\end{center}
	\end{table}
		
		APT results indicate a higher concentration of impurity elements in the Fe10Cr alloy than reported by the manufacturer (analysed by glow discharge mass spectrometry \cite{Coze2007}). For N, O and P, the concentrations measured by APT are between 7 to 11 times higher than the manufacturer-reported values. For C, the difference is a factor of 30. It is important to consider that the APT sample preparation required FIB milling and lift-out, which itself will introduce some contamination. Similar discrepancies between the manufacturer-reported values and those measured by APT have been reported previously for samples cut from the same raw material \cite{Hardie2013a}. The true amount of carbon enrichment following irradiation is probably around 100--200 appm (20--40 wppm) for the 0.8 dpa samples.
		
		It is also important to note that since the carbon contamination originated from the irradiation process, the level of contamination scales with irradiation time and dose. This means for samples irradiated to less than 0.8 dpa, the carbon enrichment is not expected to play a significant role. From a visual assessment of the materials, only samples irradiated to 0.8 dpa and above show slight to moderate browning due to carbon deposition on the surface.

\bibliographystyle{unsrt}
\bibliography{references}

\end{document}